\documentclass[final,3p,times]{elsarticle}
\makeatletter
\def\ps@pprintTitle{%
 \def\@oddfoot{\footnotesize\itshape%
   \begin{minipage}[t]{\textwidth}
   \centering Preprint \hfill \@date 
   \end{minipage}}%
 \let\@evenfoot\@oddfoot}
\makeatother

\usepackage{graphicx}
\usepackage{color,soul} 
\usepackage{array}
\usepackage{amsmath}
\usepackage{amsfonts}
\usepackage{amssymb}
\usepackage{amsthm}
\usepackage{psfrag}
\usepackage{epstopdf}
\usepackage{subcaption}
\usepackage{caption}
\usepackage{booktabs}
\usepackage{hyperref}
\usepackage{latexsym}
\usepackage{kantlipsum}
\usepackage{balance}
\usepackage{lscape}
\usepackage{rotating}
\usepackage{multicol}
\usepackage{multirow,bigdelim}
\usepackage{color, colortbl}
\usepackage{threeparttable}
\usepackage{float}
\usepackage{algorithm, algorithmicx, algpseudocode}
\usepackage{footnote}
\usepackage{cleveref}

\usepackage{xcolor}
\usepackage[author={M.}]{pdfcomment}
\definecolor{lightblue}{HTML}{ADD8E6}

\hypersetup{
    colorlinks=true,
    linkcolor=lightblue,
    urlcolor=blue,
    citecolor=lightblue
}

\newcommand{\dparder}[2]{\dfrac{\partial #1}{\partial #2}}

\newcommand{\dtotder}[2]{\dfrac{d #1}{d #2}}

\newcommand{\eval}[2][\right]{\relax\ifx#1\right\relax \left.\fi#2#1\rvert}

\newcommand{\bn}{\boldsymbol{n}}

\newcommand{\bx}{\boldsymbol{x}}

\newcommand{\bB}{\boldsymbol{B}}

\newcommand{\bN}{\boldsymbol{N}}

\newcommand{\bT}{\boldsymbol{T}}

\newcommand{\del}{\boldsymbol{\nabla}}

\newcommand{\bOmega}{\boldsymbol{\Omega}}

\makeatletter
\newcommand{\Rmnum}[1]{\expandafter\@slowromancap\romannumeral #1@}
\makeatother

\newcommand{\bmK}{\boldsymbol{\mathcal{K}}}
\newcommand{\bmM}{\boldsymbol{\mathcal{M}}}

\newcommand{\intomega}{\int_{\varOmega}}

\newcommand{\domega}{\,d\varOmega}
\newcommand{\dgamma}{\,d\varGamma}

\newcommand{\intgammah}{\int_{\varGamma_h}}

\newcommand{\varT}{T_{\delta}}
\newcommand{\bGammah}{\boldsymbol{\varGamma_h}}
\newcommand{\bGamma}{\boldsymbol{\varGamma}}

\begin{document}

\begin{frontmatter}

\title{\textcolor{black}{A finite element-based eigenvalue analysis for predicting thermal runaway in Li-ion battery packs}}
\tnotetext[t1]{}

\author[label1]{Shailendra Rahi}

\author[label2]{Vinay Dhakal}

\author[label2]{Ankur Jain}
\author[label1]{Manish Agrawal\corref{cor1}}
\ead{manish.agrawal@iitrpr.ac.in}
\cortext[cor1]{Corresponding author}


\address[label1]{Department of Mechanical Engineering, Indian Institute of Technology Ropar, Rupnagar 140001, Punjab, India}

\address[label2]{Mechanical and Aerospace Engineering Department, University of Texas at Arlington, Arlington, TX, USA}

\begin{abstract}
Thermal runaway remains a critical safety concern in battery systems and other thermally active devices, necessitating reliable methods for predicting the conditions under which thermal runaway may occur. In this work, we develop a finite element framework for assessing thermal stability in systems undergoing transient heat conduction with temperature-dependent internal heat generation. The formulation leads naturally to a generalized eigenvalue problem, wherein the sign of the smallest eigenvalue provides a direct criterion for determining the onset of thermal runaway.  This approach enables direct prediction of thermal runaway thresholds in geometrically complicated problems without requiring computationally expensive transient analysis. The proposed methodology is validated through comparison with analytical solutions for a cylindrical Li-ion cell, as well as with a pack of cylindrical cells, demonstrating excellent agreement in both cases. Based on this model, the influence of material properties, boundary conditions, geometric parameters, and spatially varying heat generation on stability limits is examined. A key advantage of this formulation is its effectiveness even for complex geometries where analytical methods become impractical. By providing a systematic and computationally efficient means to identify stability thresholds, the present work offers a practical tool for the thermal design and safety assessment of battery systems.

\color{black}

\end{abstract}


\end{frontmatter}

\section{Introduction}\label{Introduction}
The thermal safety of Li-ion cells and battery packs is of paramount concern in electrochemical energy storage in a variety of engineering systems, including electric vehicles, grid energy storage and aerospace vehicles \cite{hasan2021review,goodenough2014electrochemical}. The overheating of Li-ion batteries results in severe safety concerns, including catastrophic failure due to thermal runaway, which refers to a positive feedback between temperature-dependent decomposition reactions and temperature rise due to heat generated by such reactions \cite{bandhauer2011critical,mishra2023thermal}. Elaborate thermal management systems are often utilized to reduce temperature rise in Li-ion cells and battery packs. Energy storage devices are also often over-designed in order to minimize the risk of thermal runaway.

A fundamental understanding of reaction kinetics \cite{spotnitz2003abuse} and thermal transport \cite{bandhauer2011critical,mishra2023thermal} is of critical importance for understanding thermal runaway in Li-ion cells and to proactively predict and prevent the onset and propagation of thermal runaway. A particularly important question of practical interest is to determine under what conditions does a Li-ion cell or battery pack become thermally unstable when subjected to thermal abuse such as elevated temperature \cite{mishra2023thermal,mishra2021multi}. While stability analysis has been used extensively in fields such as fluid mechanics \cite{ding1999three,crouch2007predicting} and mechanical vibrations \cite{villa2008stability}, its use in thermal systems has been rather limited. In the specific context of Li-ion cells, thermal stability analysis has primarily been carried out using analytical approaches, including pole analysis \cite{jain2023analysis,yang2009thermal} and analysis of the nature of eigenvalues that appear in the eigenfunction-based analytical solutions of the transient temperature distribution \cite{jain2021imaginary,krishnan2022derivation,shah2016experimental}. Such analytical studies have been used to predict the conditions in which a homogeneous \cite{shah2016experimental} or multilayer Li-ion cell \cite{esho2018measurements, jain2021imaginary} is expected to remain thermally stable. The scenario of immersion cooling of a Li-ion cell has also been analyzed using pole analysis \cite{jain2023analysis}. Thermal stability analysis of other systems such as power MOSFETs \cite{narasimhan2024theoretical} and chemical reactors \cite{frank2015diffusion} is also available in the literature.

A key limitation of these analytical approaches for thermal stability prediction is their limited capability to analyze geometrically complex systems. Most analytical techniques can be applied only to simplified problems, such as a homogeneous \cite{shah2016experimental} or one-dimensional \cite{jain2021imaginary} geometry, whereas most energy storage systems have considerably complicated geometry. For example, a typical battery pack consists of multiple cylindrical cells arranged in a regular array within a rectangular box. In such a case, the inconsistency between the cylindrical shape of individual cells and the Cartesian nature of the battery pack is difficult to model using analytical tools. While a recent work \cite{dhakal2026prediction} has addressed this geometry by combining a Heaviside functions based description of the geometry with eigenfunction based stability analysis, such an approach may be computationally cumbersome and time consuming, particularly when the number of cells is large. Such an approach also fails in case of other geometrically complex features such as metal bus bars that connect cells with each other. Clearly, the development of thermal stability analysis techniques that can handle geometrically complex problems will be of much practical value, particularly for early-stage thermal design and run-time thermal control. Numerical techniques such as the finite element method provide such capability. While these methods involve approximation related to spatial discretization, however, with appropriate mesh refinement, a desirable high degree of accuracy can be achieved. 



Finite element based numerical methods have been widely used for stability analysis in several branches of engineering. Early work demonstrated the capability of these methods for predicting instability phenomena in nonlinear structural systems. For example, finite element formulations have been developed for the analysis and prediction of critical load level in geometrically nonlinear structures \cite{mallett1968finite}. Similar approaches have been applied to structural stability problems such as the lateral buckling of beams, where finite element discretization leads to eigenvalue problems for determining critical loads and corresponding instability modes \cite{attard1986lateral}. The applicability of these techniques has also been demonstrated for dynamic stability problems, where the stability of beams and framed structures has been determined numerically as a function of loading parameters and vibration frequency \cite{briseghella1998dynamic}. In geotechnical engineering, finite element approaches have been used to evaluate the stability of slopes using nonlinear failure criteria \cite{li2007finite}. Statistical finite element simulations have also been employed to investigate the stability and buckling behavior of honeycomb structures, highlighting the influence of material and geometric imperfections on the onset of instability \cite{asprone2013statistical}. 

Finite element techniques have also been employed for stability analysis in fluid and, to a lesser extent, thermal systems. For instance, linear stability analysis of incompressible viscous flows has been performed using mixed finite element formulations, where discretization of the governing equations results in generalized eigenvalue problems whose solutions yield growth rates of perturbations and critical instability parameters \cite{ding1999three}. In thermo-mechanical systems, similar formulations have been developed to study thermoelastic instability in applications such as clutches and brakes, where perturbation analysis leads to eigenvalue problems governing the exponential growth of temperature disturbances \cite{yi2000eigenvalue}. In addition to stability analysis, finite element techniques have also been used to simulate transient heat conduction and thermo-mechanical behavior in heterogeneous materials and structures \cite{mora2024high,cui2016steady,saini2021numerical,wang2005transient}, as well as for reduced-order modeling of thermal systems using modal or proper orthogonal decomposition approaches \cite{bialecki2005proper}. 

In the specific context of Li-ion batteries, a number of numerical studies have employed finite element, computational fluid dynamics (CFD), and multiphysics approaches to investigate thermal runaway and its propagation in battery cells and packs \cite{yeow2013characterizing,kariyawasam2026numerical,ZHANG2024123004}. Early three-dimensional finite element models have been developed to simulate temperature evolution under thermal abuse conditions \cite{guo2010three}, while coupled electro-thermal formulations have been used to capture heat generation and runaway triggering in battery modules \cite{chen2015thermal}. More recent studies have incorporated conjugate heat transfer and CFD-based approaches to model complex phenomena such as venting, jet fires, and thermal runaway propagation across cells \cite{kong2022coupled,wang2026numerical,li2022numerical,ZHANG2020114440}. In addition, the influence of cooling strategies and system-level design parameters on the suppression of thermal runaway propagation has been studied numerically \cite{li2022numerical}, while multiphysics and reduced-order modeling frameworks have been proposed to improve computational efficiency and scalability for large battery systems \cite{ling2024thermal,parhizi2024reduced}. 



Despite the literature cited above, the use of finite element based eigenvalue analysis techniques for directly determining thermal stability limits in Li-ion cells and battery packs has remained largely unexplored. In contrast with the extensive use of such techniques for stability analysis in structural, fluid and thermo-mechanical systems, relatively little work has been carried out for predicting the onset of thermal runaway in Li-ion cells and battery packs. In particular, most existing numerical studies on battery thermal behavior have used full-scale transient thermal simulations to predict the propagation of thermal runaway, rather than on directly determining the conditions under which thermal runaway is expected to occur. 

Given the relatively limited literature on eigenvalue based numerical techniques for thermal stability analysis, and the clear importance of predicting thermal stability of Li-ion cells and battery packs, this work presents a finite element method based technique to determine thermal response of a pack of cylindrical Li-ion cells to a thermal excursion and, in particular, to determine whether a given battery pack is going to be thermally stable or not. 
This problem is solved by first linearizing the temperature-dependent internal heat generation term about a reference state. The linearized governing heat conduction equation is then discretized using the finite element method and the resulting semi-discrete system is reformulated as a generalized eigenvalue problem. The sign of the smallest eigenvalue is then used as a criterion to determine whether the system is thermally stable or unstable. It is shown that the proposed approach accurately reproduces analytical stability results reported in the past for a single cylindrical cell as well as for the matrix consisting of multiple cells. The formulation is utilized to obtain the threshold conditions in more complex scenarios involving spatially varying heat generation, non-uniform convective boundary conditions, and variations in cell size. Finally, the capability of the formulation to analyze complex three-dimensional pack configurations is demonstrated. The proposed formulation enables systematic investigation of the influence of thermophysical properties and cooling conditions on the onset of thermal runaway, thereby providing useful insights for the thermal design and safety assessment of Li-ion battery packs.

The structure of the remainder of this manuscript is as follows. In \autoref{methodology}, we present an outline of the finite-element method based technique. This section provides a description of the governing equations and their discretization within the finite element framework, together with the eigenvalue formulation used to determine thermal stability. In \autoref{Results}, we compare the present method with related past work and discuss results for several representative configurations, including a single cylindrical cell, battery pack arrangements, and more complex scenarios involving spatially varying parameters and three-dimensional geometries. The paper concludes with a summary of the main findings.

\section{Mathematical model}\label{methodology}
\begin{figure}[ht!]
\centering
\includegraphics[trim= {0.0cm 0.0cm 5.0cm 0.0cm},clip,width=0.7\textwidth]{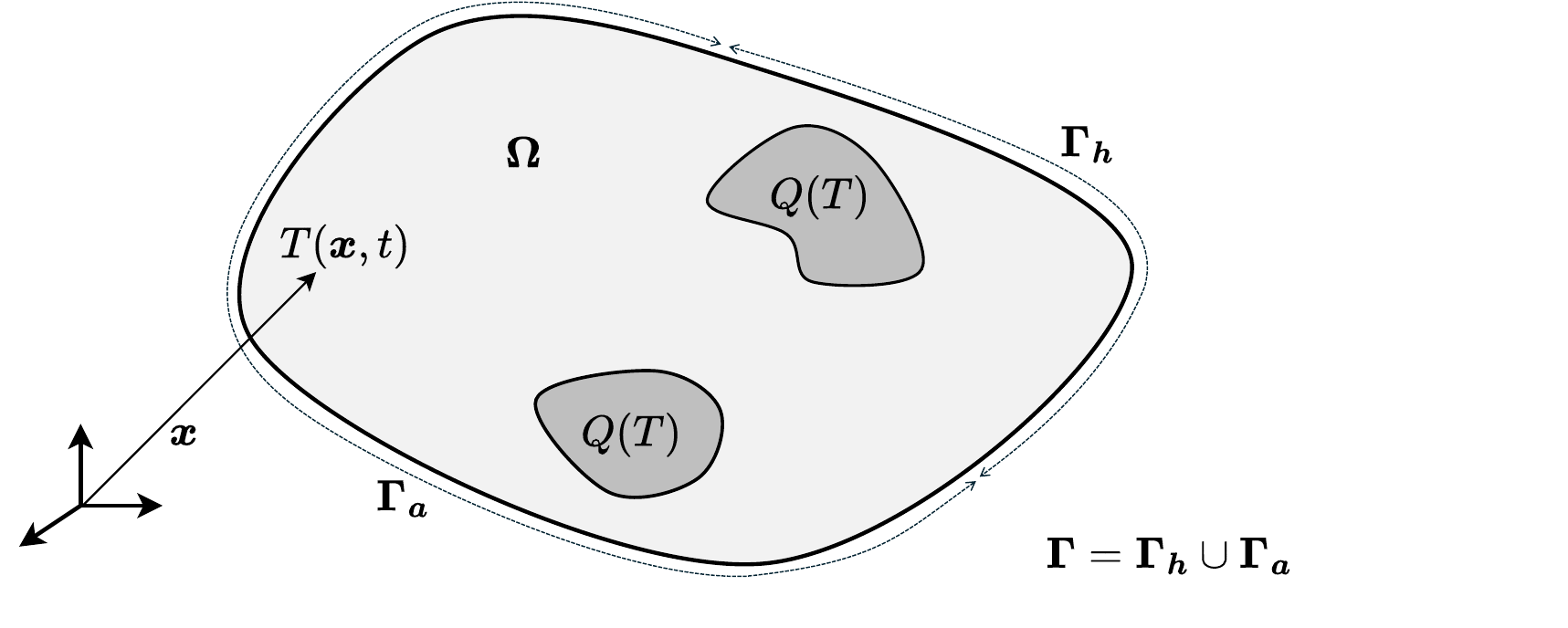}
\caption{ Schematic of the heat generating body occupying domain $\bOmega$, bounded by surface $\bGamma$, with $\bGammah$ and $\bGamma_a$ denoting the portions of the boundary subjected to convective heat transfer and adiabatic conditions, respectively. Temperature-dependent heat generation, denoted by $Q(T)$, occurs inside the body.}
\label{fst}
\end{figure}

We consider a general heat generating body occupying a domain denoted by $\bOmega$, as shown in \autoref{fst}. Heat generation may occur throughout the entire domain of the body, such as in a Li-ion cell, or within localized regions, as in a battery pack comprising multiple cells. The heat generation rate in the body is assumed to be dependent on its temperature, so that an increase in temperature may lead to an increase in heat generation. An initial positive temperature field may cause a positive feedback loop, which is referred to as thermal runaway, and is of primary interest in this work. The domain is bounded by the surface denoted by $\bGamma$, over which, the body loses heat by convective cooling. The objective is to develop a numerical framework for predicting thermal runaway of this body by examining thermal stability of the system undergoing transient heat transfer with temperature-dependent internal heat generation. The temperature field within the body evolves due to transfer of internally generated heat, first through the body due to conduction and then into the ambient through convective heat transfer at the surface of the body. Heat exchange with the ambient is modeled through a convective boundary condition imposed on $\bGammah \subseteq \bGamma$. The spatial variable in the domain is denoted by $\bx$. Temperature at a space point $\bx \in \bOmega$ and time $t$ is denoted by $T(\bx,t)$. 

\subsection{Development of thermal model for thermal runaway scenarios}\label{tdm}
The transient heat conduction equation governing the thermal response of the system is given by \cite{ozicsik1993heat}
\begin{equation}
    \del \cdot (k \del T) + Q(T) = \rho c \dparder{T}{t}, \ \ \ \ \ \text{in} \ \bOmega,
    \label{heat_eq}
\end{equation}
subject to the convective boundary condition
\begin{equation}
    -k \del T \cdot \bn  = h T, \ \ \ \ \ \text{on} \ \bGammah,
    \label{cbc}
\end{equation}
and the initial condition
\begin{equation}
    T(\bx,t=0)  = T_i(\bx),
    \label{inc}
\end{equation}
where $\rho(\bx)$ is the density, $c(\bx)$ is the specific heat capacity, $k(\bx)$ is the thermal conductivity that may be anisotropic in general 
, $\bn$ denotes the outward unit normal to the surface $\bGamma$, $h(\bx)$ is the convective heat transfer coefficient on the boundary, $T_i(\bx)$ is the initial temperature profile and $Q(T)$ represents the volumetric heat generation that depends on temperature.

The nature of the solution to equations \eqref{heat_eq}, \eqref{cbc} and \eqref{inc} determines the conditions under which thermal runaway may occur. The existence of a bounded solution at large time implies that heat generated within the body is sufficiently dissipated through conduction and convection at boundaries, whereas an unbounded solution signifies thermal instability. Therefore, it is of practical interest to determine the design space in which thermal stability may be expected, as well as the surface in the design space that separates thermally stable and unstable regions. While equation \eqref{heat_eq} is difficult to solve for a general $Q(T)$, which is often strongly non-linear in nature due to the Arrhenius kinetics of decomposition reactions that cause heat generation in a Li-ion cell \cite{spotnitz2003abuse}, we adopt a linearization approach to enable the development of a numerical framework. Towards this objective, we carry out a Taylor series expansion of the heat generation term $Q(T)$ about a temperature $T_0(\bx)$, and neglect second and higher order terms. This reduces equation \eqref{heat_eq} to

\begin{equation}
    \del \cdot (k \del T) + Q(T_0) + \beta(T-T_0) = \rho c \dparder{T}{t} \ \ \ \ \ \text{in} \ \bOmega,
    \label{lheat_eq}
\end{equation}
where $\beta(\bx) = \left.\dtotder{Q}{T}\right|_{T_0(\bx)}$ represents the slope of $Q(T)$.

By doing so, the heat generation term can now be expressed as the sum of two terms: ($Q(T_0)-\beta T_0$) and $\beta T$. The first term is a constant volumetric heat source, which leads to a bounded steady-state temperature distribution, provided the boundaries are not completely adiabatic. Since thermal runaway is associated with the loss of boundedness of the temperature field rather than its magnitude, this term does not contribute to thermal instability. On the other hand, the second term increases linearly with temperature and therefore may introduce a positive thermal feedback into the system, wherein an increase in temperature results in greater internal heat generation, which further elevates the temperature, and so on. Under certain conditions \cite{jain2021imaginary,krishnan2022derivation}, this self-amplifying feedback mechanism may cause the temperature to grow without bound, ultimately triggering thermal runaway.

Having identified the fundamentally different roles of the two heat generation components in equation \eqref{lheat_eq}, it follows that only the temperature-dependent term, i.e. $\beta T$, has the potential to induce thermal instability. Accordingly, stability analysis may be carried out by focusing exclusively on this linearly temperature-dependent heat generation term. This is equivalent to examining the evolution of a temperature perturbation about a reference state, wherein constant contributions vanish and only terms capable of amplifying the thermal disturbance are retained. Accordingly, for the purpose of stability analysis, we retain only the linearly temperature-dependent heat generation term, and rewrite equation \eqref{lheat_eq} as

\begin{equation}
    \del \cdot (k \del T) + \beta T = \rho c \dparder{T}{t} \ \ \ \ \ \text{in} \ \bOmega,
    \label{lheat2_eq}
\end{equation}
subject to the same boundary condition as given by equation \eqref{cbc}. Since the present formulation is based on a linearization of the governing heat generation term about a temperature $T_0(x)$, the resulting system is strictly valid for small perturbations in the vicinity of this state. However, the objective of the analysis is not to capture the full nonlinear thermal evolution, but rather to determine the onset of thermal instability. In this context, the linearized system provides a criterion for stability at the given instant. Specifically, if a perturbation introduced at $T_0(\bx)$ exhibits exponential growth, the system becomes unstable, and the temperature will continue to increase under the full nonlinear dynamics. Thus, the proposed linearization approach enables the identification of the critical conditions for the onset of thermal runaway without requiring nonlinear transient analysis.

The governing equations \eqref{cbc} and \eqref{lheat2_eq} form the foundation for the finite element formulation developed in this work, enabling a rigorous assessment of whether temperature disturbances decay or grow with time, and thereby providing a direct criterion to predict whether thermal runaway will occur or not.

\subsection{Variational formulation}\label{varf}
Having established the governing equations that characterize the thermal system, we now proceed to develop the variational framework that forms the basis of the finite element formulation. The weak form of the governing equation is obtained by multiplying equation \eqref{lheat2_eq} by an admissible temperature variation $\varT$ and integrating over the domain $\bOmega$. Applying the divergence theorem to the diffusion term yields

\begin{equation}
    \intomega \rho c \dparder{T}{t} \varT \domega + \intomega \del T \cdot k \del \varT \domega + \intgammah h T \varT \dgamma - \intomega \beta T \varT \domega = 0, \ \ \ \ \ \forall \ \varT \in \mathcal{V},
    \label{weakf}
\end{equation}
where $\mathcal{V}$ denotes the space of admissible temperature variations satisfying the essential boundary conditions.

\subsection{Finite element discretization}\label{fem}
To construct the finite element approximation, the temperature field and its variation are interpolated using the standard discretizations
\begin{equation}
\begin{aligned}
    T(\bx,t) = \bN(\bx) \tilde{T}(t), \\
    \varT(\bx,t) = \bN(\bx) \tilde{T}_\delta(t),
    \label{ipT}
\end{aligned}
\end{equation}
where $\bN(\bx)$ is the matrix of shape functions and $\tilde{T}(t)$ represents the vector of nodal temperatures.

Using the interpolated form, we have


\begin{equation}
\begin{aligned}
    \del T(\bx,t) = \bB(\bx) \tilde{T}(t), \\
    \del \varT(\bx,t) = \bB(\bx) \tilde{T}_\delta(t), \\
    \dparder{T}{t} = \bN(\bx) \dot{\tilde{T}}(t),
    \label{ipdT}
\end{aligned}
\end{equation}
where 
the $(i,j)$ th element of $\bB(\bx)$ matrix is given by $N_{i,j}=\dparder{N_i}{x_j}$, representing the partial derivative of $i$-th shape function with respect to $j$-th spatial coordinate.
\newline

Substituting the above quantities and exploiting the arbitrariness of $\varT$, the matrix form of the equation \eqref{weakf} may be obtained as



\begin{equation}
    \bmM \dot{\tilde{T}}(t) + \bmK \tilde{T}(t)  = 0,
    \label{mateq}
\end{equation}
where the mass matrix $\bmM$ and the effective conductivity matrix $\bmK$ are given by
\newline

$
\begin{aligned}
    \bmM &= \intomega \rho c \bN^{\bT} \bN \domega, \\
    \bmK &= \intomega  k \bB^{\bT} \bB \domega -
           \intomega \beta \bN^{\bT} \bN \domega +
           \intgammah h \bN^{\bT} \bN \dgamma.
    \label{massmat}
\end{aligned}
$
\newline

It is evident that the contribution of the temperature-dependent heat generation appears in the effective conductivity matrix, thereby directly influencing the stability characteristics of the resulting semi-discrete system. This system constitutes a linear dynamical problem whose temporal behavior is governed by the properties of the mass and the conductivity matrices, both of which encapsulate the thermo-physical properties of the problem. Consequently, the onset of thermal instability may be characterized by examining whether the system admits exponentially growing temperature modes, a question that is naturally addressed by an eigenvalue analysis.

\subsection{\textcolor{black}{Generalized eigenvalue problem}}
The semi-discrete finite element equations obtained in the previous section constitute a linear system of ordinary differential equations in time. Owing to this structure, the system naturally admits solutions whose temporal evolution may be expressed in exponential form. Accordingly, we seek solution of equation \eqref{mateq} in the form
\begin{equation}
    \tilde{T}(t) = \hat T \exp({-\lambda t}),
    \label{exp}
\end{equation}
where $\lambda$ represents the temporal growth parameter.

By substituting the exponential representation of the temperature field in equation \eqref{mateq}, we obtain a generalized eigenvalue problem, given by
\begin{equation}
     \bmK \hat T = \lambda \bmM \hat T,
    \label{egv}
\end{equation}
where $\lambda$ are the eigenvalues and $\hat T$ are the associated eigenmodes. Since the matrices $\bmM$ and $\bmK$ are symmetric, all eigenvalues are real.

Each eigenvalue corresponds to a distinct temperature mode whose temporal behavior is governed by its sign. In view of the exponential representation of the temperature field introduced in equation \eqref{exp}, positive eigenvalues lead to exponentially decaying temperature modes, ensuring that the transient response of the system governed by equation \eqref{lheat2_eq} remains bounded. Under such conditions, thermal runaway does not occur. In contrast, negative eigenvalues give rise to exponentially growing temperature modes, indicating a loss of boundedness and signaling the onset of thermal runaway. The thermal response therefore remains stable only when all eigenvalues are positive. The presence of even a single negative eigenvalue is sufficient to trigger thermal runaway, as the associated temperature mode amplifies continuously and drives the temperature towards unbounded growth.

\subsection{\textcolor{black}{Criterion for thermal runaway}}
The generalized eigenvalue problem given in previous section provides a direct means for identifying the stability limit of the thermal system governed by equation \eqref{lheat2_eq}. Eigenvalue-based stability analysis is widely used across engineering disciplines \cite{crandall1973bifurcation,makarov1998general,somieski2001eigenvalue,luongo2000sensitivities}, where the onset of instability is characterized by the vanishing of the smallest eigenvalue of the governing system. The present work adopts this criterion within the context of thermal stability of heat-generating systems. Since the temporal evolution of each temperature mode is governed by the corresponding eigenvalue, the transition from stable to unstable behavior occurs when a decaying mode becomes neutrally stable and subsequently begins to grow. This critical condition is reached when the smallest eigenvalue of the system approaches zero. 

Accordingly, the threshold for thermal runaway is characterized by the condition
\begin{equation}
    \lambda_{min} = 0,
\end{equation}
which represents the point at which the system loses thermal stability. For $\lambda_{min} > 0$, all admissible temperature modes decay with time and transient response remains bounded. Conversely, when $\lambda_{min} < 0$, at least one mode grows exponentially, indicating the onset of thermal runaway. 

Having established the eigenvalue-based criterion for predicting thermal runaway, we now employ the proposed finite element framework to investigate the onset of thermal runaway in representative thermal systems. The formulation enables a systematic evaluation of the stability characteristics by directly computing the eigenvalues associated with the discretized governing equations.

In the following section, we present a series of numerical examples that validate the proposed formulation and demonstrate its predictive capability.


\section{Results and discussion}\label{Results}
In this section, we demonstrate the capability of the proposed finite-element framework to predict thermal runaway in Li-ion cells and battery packs through a sequence of example problems of increasing complexity. These examples are designed to (1) validate the formulation against analytical results for special cases available in the literature, (2) examine its ability to accurately identify the onset of thermal instability, and (3) demonstrate its applicability to geometrically complicated configurations that closely resemble practical battery systems and are difficult to model using available analytical techniques. 

We begin with a single cylindrical cell problem for which an analytical solution exists \cite{shah2016experimental}, thereby providing a direct means of verification. The study is then extended to a battery pack model to investigate the influence of material properties, boundary conditions, and spatial variations in heat generation on thermal stability. Subsequently, more realistic scenarios involving non-uniform cooling, heterogeneous heat generation, and geometric variations are considered to highlight the robustness and versatility of the framework. Finally, the formulation is applied to three-dimensional problems to demonstrate that the approach readily generalizes beyond two-dimensional analyses. Collectively, these examples establish the proposed method as a reliable and versatile numerical tool for assessing thermal stability of Li-ion cells and battery packs, as well as other related systems in which temperature-dependent heat generation occurs.

In the presented simulations, we employ 9-node and 27-node conventional finite elements for two-dimensional and three-dimensional problems respectively. For all simulations, a mesh convergence study is performed using progressively refined meshes, and grid independence is confirmed.

\subsection{Cylindrical cell problem}
In order to validate the proposed framework, we first consider the thermal stability of a single cylindrical Li-ion cell, which has been addressed analytically in \cite{shah2016experimental}. The availability of an analytical solution allows us to directly assess the accuracy of the numerical predictions.

\begin{figure}[ht!]
\centering
\includegraphics[trim= {0.0cm 0.0cm 0.0cm 0.3cm},clip,width=0.25\textwidth]{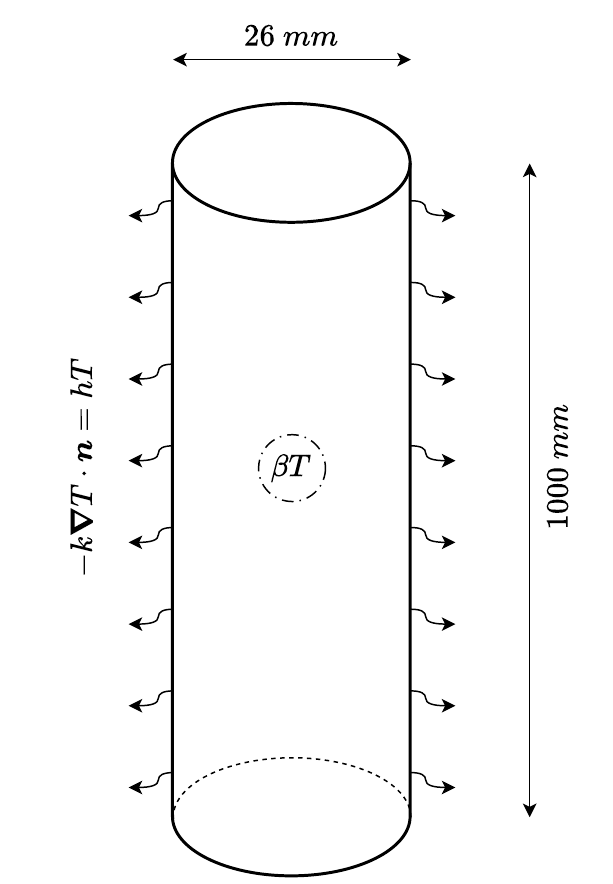}
\caption{Schematic of the cylindrical cell. The cell is modeled as an infinitely long cylinder by taking its length sufficiently large compared to its diameter. The cell is subjected to internal temperature-dependent heat generation while the convective heat transfer with coefficient $h$ is prescribed on the outer surface.}
\label{cyl}
\end{figure}

An infinitely long cylindrical cell subjected to internal heat generation is considered, with the heat generation parameter taken as $\beta = 6000\ Wm^{-3}K^{-1}$. The geometry of a standard 26650 Li-ion cell (diameter of $26\ mm$) is considered, as shown in \autoref{cyl}. The outer surface of the cell is subjected to convection with a uniform heat transfer coefficient $h$, and a uniform radial thermal conductivity $k$ is assumed throughout the domain. Even though thermal conductivity of a Li-ion cell is known to be anisotropic, the problem being considered is spatially one-dimensional, and, therefore, only the radial thermal conductivity needs to be considered. While $k$ governs heat conduction within the cell, $h$ represents the extent of heat dissipation from the surface.

\begin{figure}[ht!]
\centering
\includegraphics[trim= {0.0cm 0.0cm 8.0cm 0.0cm},clip,width=0.75\textwidth]{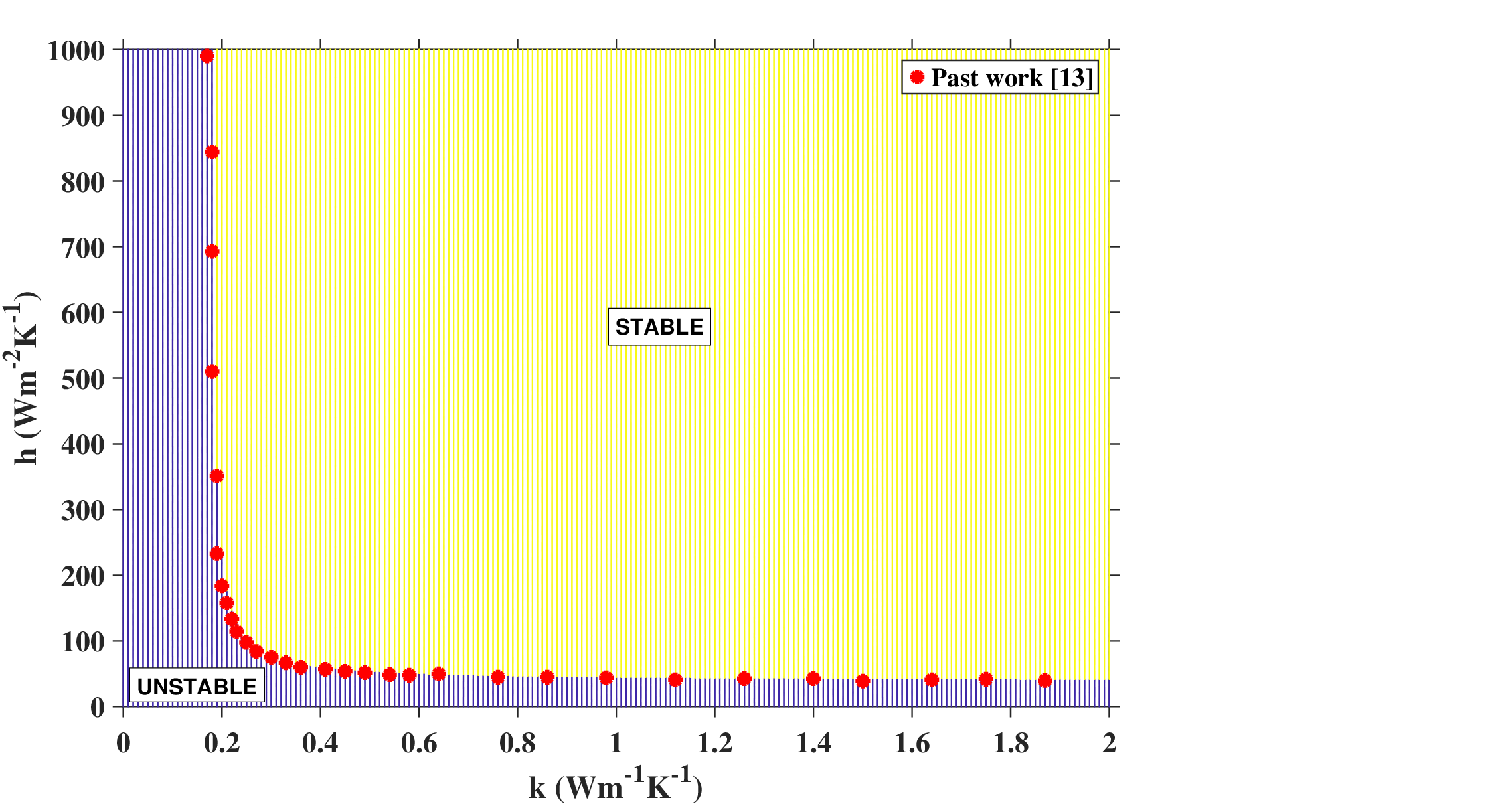}
\caption{Scatter plot identifying stable and unstable regions in the $k-h$ design space. The two distinct regions correspond to the states of thermal stability and thermal runaway which are determined by the sign of the smallest eigenvalue obtained for a given $(k,h)$ pair. For this analysis, $\beta=6000\ Wm^{-3}K^{-1}$ is prescribed. Analytical transition points from \cite{shah2016experimental}, marked as discrete markers, are superimposed for comparison, demonstrating excellent agreement with the numerical results.}
\label{scatterr}
\end{figure}

Since both the geometry and boundary conditions are axi-symmetric, the problem is modeled using axi-symmetric finite elements, with a total of $20$ elements. The thermal stability of the cell is analyzed by varying the parameters $k$ and $h$, and examining the sign of the smallest eigenvalue of the resulting system. For any pair $(k,h)$, smallest eigenvalue being positive indicates a stable thermal response, whereas a negative value indicates thermal runaway. To systematically study the effect of these governing thermal parameters, simulations are carried out for conductivity values in the range $[0,2]\ Wm^{-1}K^{-1}$ and heat transfer coefficients in the range $[0,1000]\ Wm^{-2}K^{-1}$, with $\beta=6000\ Wm^{-3}K^{-1}$. The resulting scatter plot in the $k$--$h$ space is shown in \autoref{scatterr}. The plot contains two distinct regions corresponding to stable and unstable operating conditions, thereby identifying the safe and unsafe portions of the design space. For comparison, points representing stability thresholds--i.e., the minimum value of $h$ required for stability at a given $k$--based on a past analytical study of this problem \cite{shah2016experimental} are also marked.

The numerical predictions are found to be in very close agreement with the analytical results reported in \cite{shah2016experimental}. The threshold points based on past work lie almost exactly on the curve that is found to separate the stable and unstable regions identified by the present work. This excellent agreement verifies the accuracy of the proposed formulation for this specific problem, and demonstrates its ability to accurately predict the onset of thermal runaway in cylindrical cells. The resulting stability map also provides useful physical insight into the role of thermal properties and external cooling. In particular, high thermal conductivity is found to promote thermal stability by facilitating more efficient redistribution of internally generated heat within the cell, thereby reducing the external cooling required to maintain stability. Conversely, cells with low thermal conductivity are found to require much stronger external convective cooling to prevent the onset of thermal runaway.

\subsection{Battery pack problem}
\label{secbase}
\begin{figure}[ht!]
\centering
\includegraphics[trim= {0.0cm 0.0cm 0.0cm 0.0cm},clip,width=0.99\textwidth]{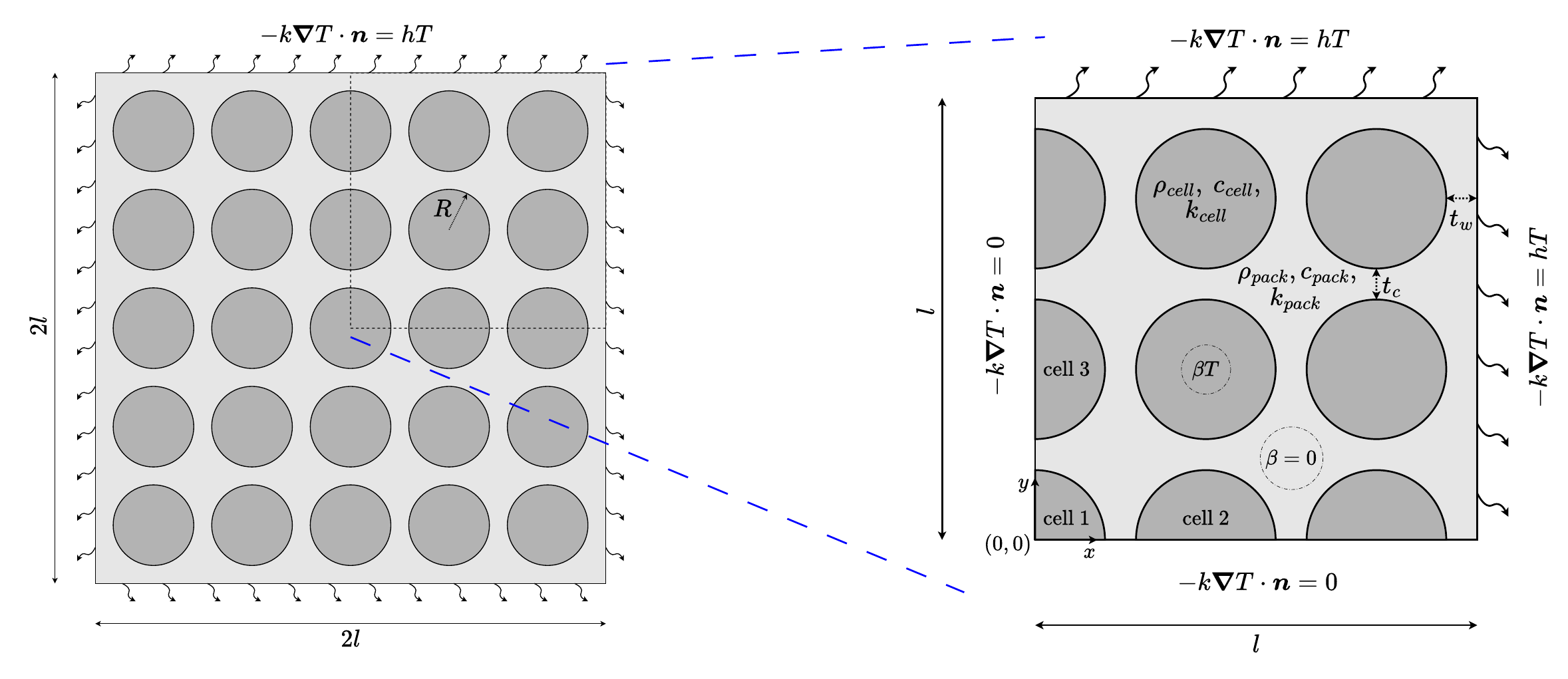}
\caption{Schematic of the battery pack comprising 25 cylindrical cells arranged in a 5 by 5 configuration within a square pack domain. Heat generation occurs only within the cells, while all the outer boundaries are subjected to convective heat transfer. Owing to symmetry, only one quarter of the geometry is simulated, as shown in the enlarged view, with appropriate symmetry and convective boundary conditions.}
\label{fpack}
\end{figure}

In this sub-section, we investigate the thermal stability of a battery pack containing multiple cylindrical cells embedded within a interstitial pack material. The pack is modeled as a two-dimensional square domain containing 25 cylindrical cells arranged in a 5 by 5 configuration. Internal heat generation is assumed to occur only within the cells, and not in the interstitial pack material. All four edges of the pack are subjected to convective cooling. Owing to the geometrical and thermal symmetry of the problem, only one quarter of the domain is analyzed with appropriate boundary conditions, as shown in \autoref{fpack}. Convection is applied on the two outer edges, whereas the remaining two symmetry edges are treated as adiabatic. The quarter battery pack is simulated with a total of $2479$ finite elements.

\begin{table}[ht!]
\centering
\caption{Thermophysical properties used for the cell and pack material \cite{dhakal2026prediction}.}
\label{table}
\begin{tabular}{ll}
\hline
Property & Value \\
\hline
Pack dimension ($l$)  & 57 $mm$ \\
Cell radius ($R$)  & 9 $mm$ \\
Cell-to-cell gap ($t_c$) & 4 $mm$ \\
Cell-to-wall gap ($t_w$) & 4 $mm$ \\
Cell density ($\rho_{cell}$) & 2280 $kgm^{-3}$ \\
Pack density ($\rho_{pack}$) & 160 $kgm^{-3}$ \\
Cell specific heat capacity ($c_{cell}$) & 715 $Jkg^{-1}K^{-1}$ \\
Pack specific heat capacity ($c_{pack}$) & 1562 $Jkg^{-1}K^{-1}$ \\
Cell conductivity ($k_{cell}$) & 0.2 $Wm^{-1}K^{-1}$ \\

\hline
\end{tabular}
\end{table}

\begin{figure}[ht!]
\centering
\includegraphics[trim= {0.0cm 0.0cm 8.0cm 0.0cm},clip,width=0.65\textwidth]{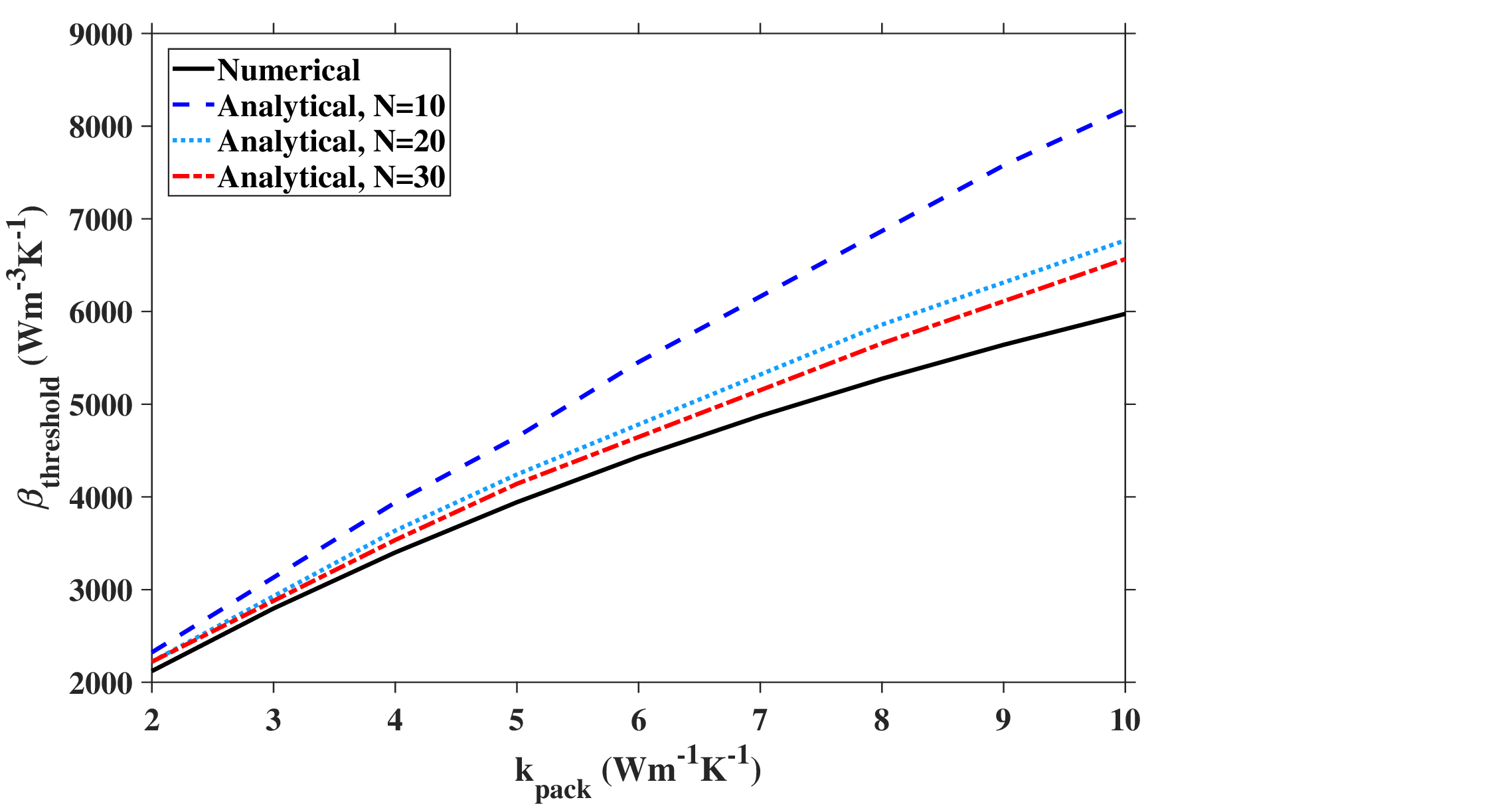}
\caption{Stability curves showing the threshold values $\beta_{threshold}$ as a function of pack thermal conductivity $k_{pack}$. The numerical prediction obtained from the proposed eigenvalue--based formulation is compared with three analytical curves derived using truncated series expansions \cite{dhakal2026prediction}. Increasing the number of terms ($N$) in the analytical solution improves agreement with the numerical result, demonstrating convergence and validating the proposed framework.}
\label{stability}
\end{figure}

We aim to determine the maximum allowable heat generation parameter $\beta$ in each cell for a given thermal conductivity of the pack material such that thermal runaway is avoided. Accordingly, we evaluate the threshold values of $\beta$ over a range of pack thermal conductivities by examining the sign of the smallest eigenvalue of the resulting system for a given pair $(k_{pack},\beta)$. A uniform convective heat transfer coefficient $h=1000\ Wm^{-2}K^{-1}$ is prescribed on the convective boundaries. The other thermophysical properties used for the cell and pack materials are summarized in \autoref{table} \cite{dhakal2026prediction}. To assess the accuracy of the proposed formulation, the numerical results are compared with analytical predictions reported in \cite{dhakal2026prediction} for the same configuration. Both numerical and analytical curves are presented together in \autoref{stability}.


The figure contains four curves: one obtained using the proposed numerical framework and the other three derived from the analytical solution with number of terms $(N)$ up to which the series solution is truncated \cite{dhakal2026prediction}. The analytical approach relies on a truncated series expansion, whose accuracy improves as additional terms are retained, but at considerably increased computational expense because the computational time scales as the fourth power of the number of terms considered. This makes it rather difficult to use the analytical technique with a large number of terms. The $N=10$ analytical curve exhibits a noticeable deviation from the numerical prediction. Upon increasing the number of terms in the series, the subsequent analytical curves shift closer to the numerical stability curve. This progressive convergence of the analytical solution toward the numerical result provides further verification of the proposed framework. The computational time for the FEM technique is several seconds compared to at least several tens of minutes for the analytical approach, demonstrating superiority of the numerical technique in terms of computational cost, particularly for geometrically complicated problems.

\begin{figure}[ht!]
\centering
\includegraphics[trim= {0.0cm 0.0cm 8.0cm 0.0cm},clip,width=0.65\textwidth]{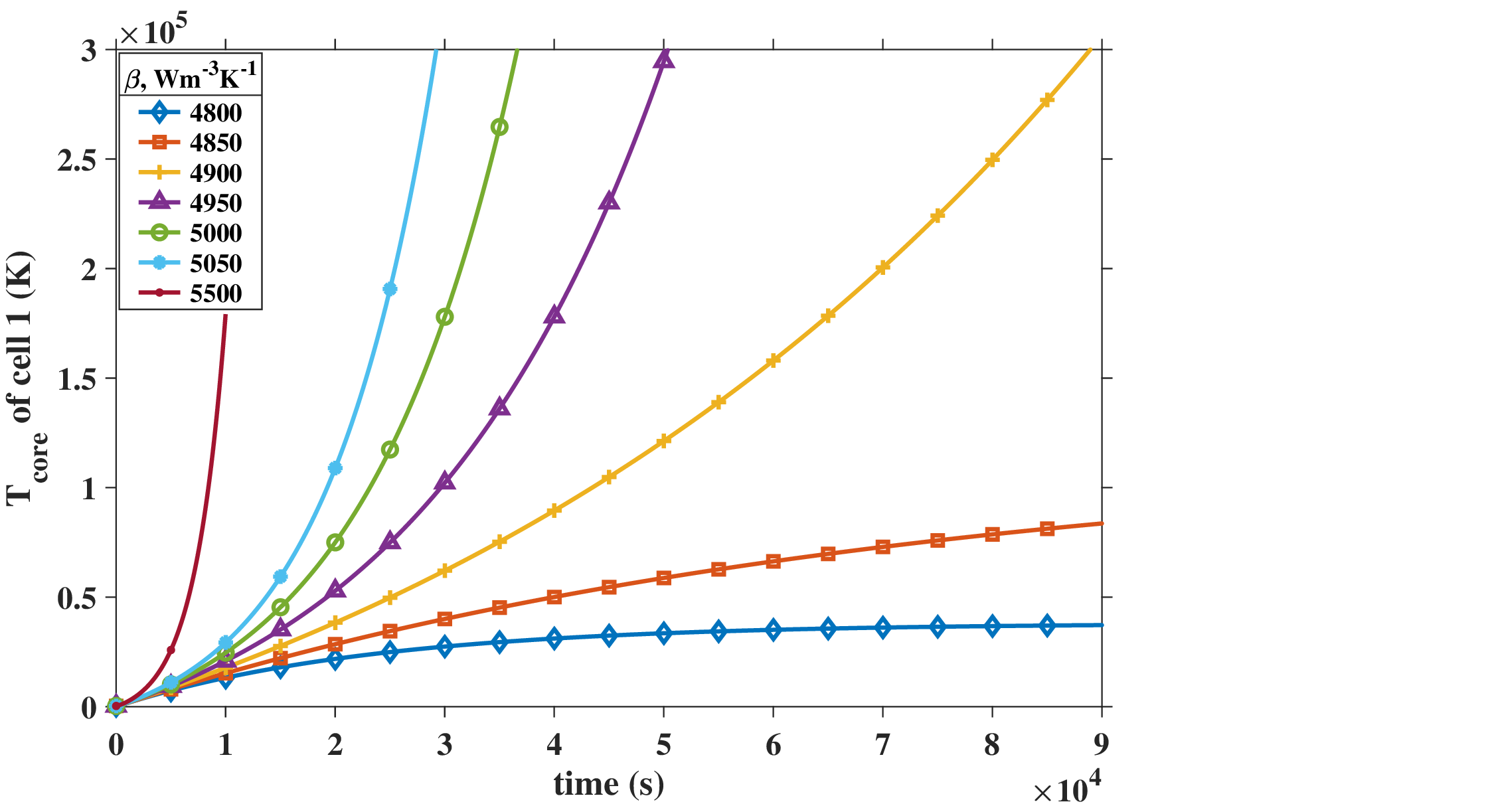}
\caption{Transient temperature responses at the location $(0,0)$ inside Cell 1 of the battery pack. The responses correspond to $k_{pack} = 7\ Wm^{-1}K^{-1}$ and $\beta = 4800, 4850\ Wm^{-3}K^{-1}$ ($< \beta_{threshold}$); $\beta=4900, 4950, 5000, 5050, 5500\ Wm^{-3}K^{-1}$ ($> \beta_{threshold}$).}
\label{tr}
\end{figure}

To further verify the predictive capability of the formulation, a transient finite element analysis is performed for a representative pack thermal conductivity of $k_{pack} = 7\ Wm^{-1}K^{-1}$. Based on the numerically obtained stability curve shown in \autoref{stability}, threshold value of $\beta$ for $k_{pack} = 7\ Wm^{-1}K^{-1}$ is $\beta_{threshold} = 4874\ Wm^{-3}K^{-1}$. Two values of $\beta$ below the threshold ($\beta = 4800, 4850\ Wm^{-3}K^{-1}$) and five values above the threshold ($\beta =$ 4900, 4950, 5000, 5050, 5500 $Wm^{-3}K^{-1}$) are considered. The temporal evolution of temperature at the center of Cell 1 is shown in \autoref{tr}.

These transient responses are fully consistent with the stability predictions. Temperature histories corresponding to heat generation values above the threshold exhibit unbounded growth, indicating the onset of thermal runaway. In contrast, temperatures associated with subcritical values of $\beta$ gradually stabilize and converge to bounded steady states. This agreement independently confirms the accuracy of thermal stability threshold prediction using the proposed approach that combines finite element method and eigenvalue analysis.

\begin{figure}[ht!]
 \centering
     \begin{subfigure}[b]{0.46\textwidth}
         \centering
         \includegraphics[trim= {0.0cm 0.0cm 5.0cm 0.0cm},clip,width=1\textwidth]{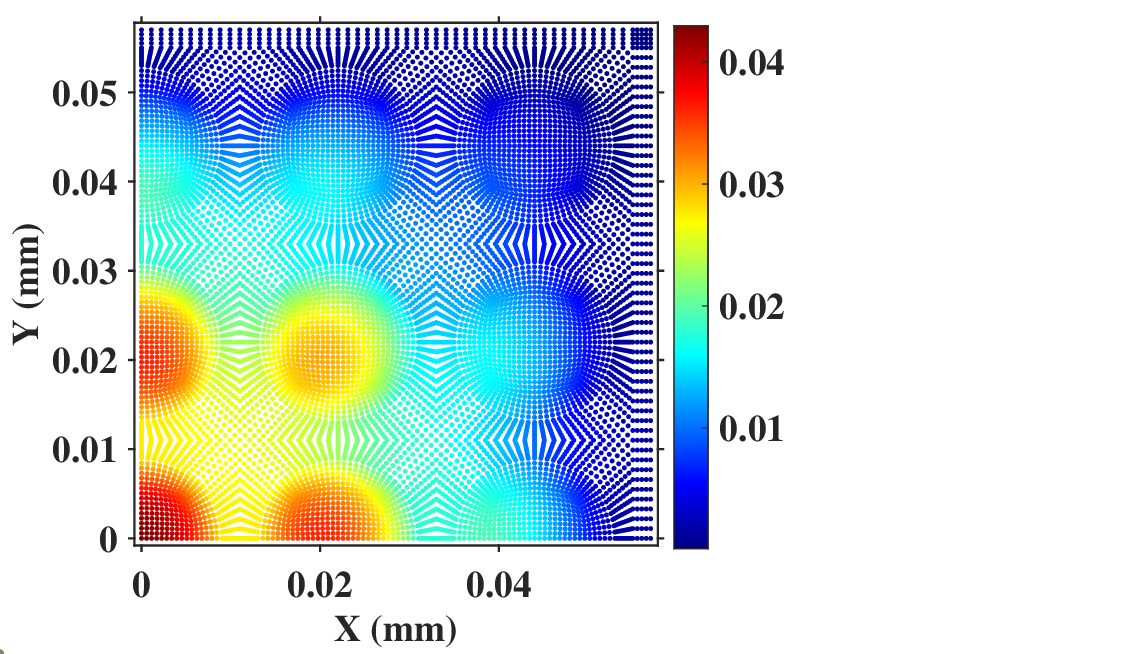}
         \caption{Unstable mode}
         \label{um}
     \end{subfigure}
     \begin{subfigure}[b]{0.46\textwidth}
         \centering
         \includegraphics[trim= {0.0cm 0.0cm 5.0cm 0.0cm},clip,width=1\textwidth]{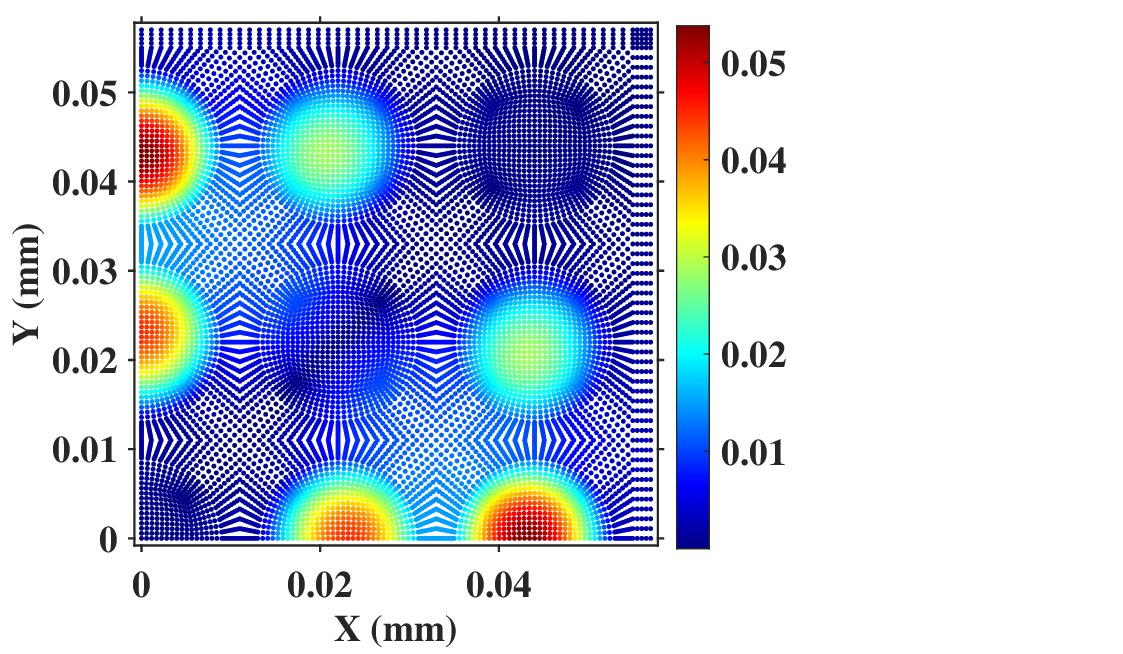}
         \caption{Stable mode}
         \label{sm}
     \end{subfigure}
 \caption{\textcolor{black}{The two eigen modes associated with a negative eigen value (\autoref{um}) and a positive eigen value (\autoref{sm}). The spatial structure of the modes highlights the regions prone to high temperatures.}}
 \label{egm}
 \end{figure}

Additional insight into the spatial characteristics of the instability is obtained by examining two representative eigenmodes that correspond to two smallest eigenvalues resulting from $k_{pack} = 7\ Wm^{-1}K^{-1}$ and $\beta = 4875\ Wm^{-3}K^{-1}$, which is just above the threshold value. The two modes are shown in \autoref{egm}. One eigenmode, shown in \autoref{um}, corresponds to the smallest eigenvalue, which is negative and therefore unstable. The stable mode, shown in \autoref{sm}, corresponds to the other eigenvalue, which is greater than the smallest and is positive. The unstable eigenmode reveals a pronounced temperature concentration within and around Cell 1, which is located near the adiabatic boundaries and therefore experiences limited heat dissipation. Conversely, the stable eigenmode shows elevated temperature near cells adjacent to the convective boundaries, where enhanced cooling promotes stability.

The observed eigenmode patterns are physically intuitive and further reinforce the robustness of the proposed methodology in capturing both the onset and spatial manifestation of thermal instability in battery packs. In particular, the unstable temperature modes tend to localize in regions where heat removal is relatively weaker. Such spatial concentration of the eigenmodes indicates that thermal runaway is more likely to initiate in these regions and subsequently propagate through the pack via conductive heat transfer between adjacent cells. These patterns, therefore, provide useful insight into the locations that are most susceptible to thermal instability and highlight the importance of effective heat removal from those regions.

It is noted that the threshold values of $\beta$ determined using the procedure described in this section can be directly related to corresponding critical temperature conditions, since $\beta$ represents the rate of increase in internal heat generation with temperature. Thus, the computed stability limits may also be interpreted in terms of the critical temperature at which thermal runaway is initiated.

\subsection{Extended battery pack simulations}

To further demonstrate the versatility of the proposed framework, we consider several more complicated battery pack heat transfer problems. The objective is to assess the predictive capability of the numerical formulation under more practical conditions, and, specifically, to develop physical insights into how spatial variations in thermal parameters influence the onset of thermal runaway. These simulations represent increasingly complex and realistic operating conditions for which analytical approaches become difficult or impractical to apply. All simulations are carried out with the same mesh as described in \autoref{secbase}, consisting of $2479$ finite elements.

\subsubsection{Spatially varying convective cooling}

\begin{figure}[ht!]
\centering
\includegraphics[trim= {0.0cm 0.0cm 0.0cm 0.0cm},clip,width=0.5\textwidth]{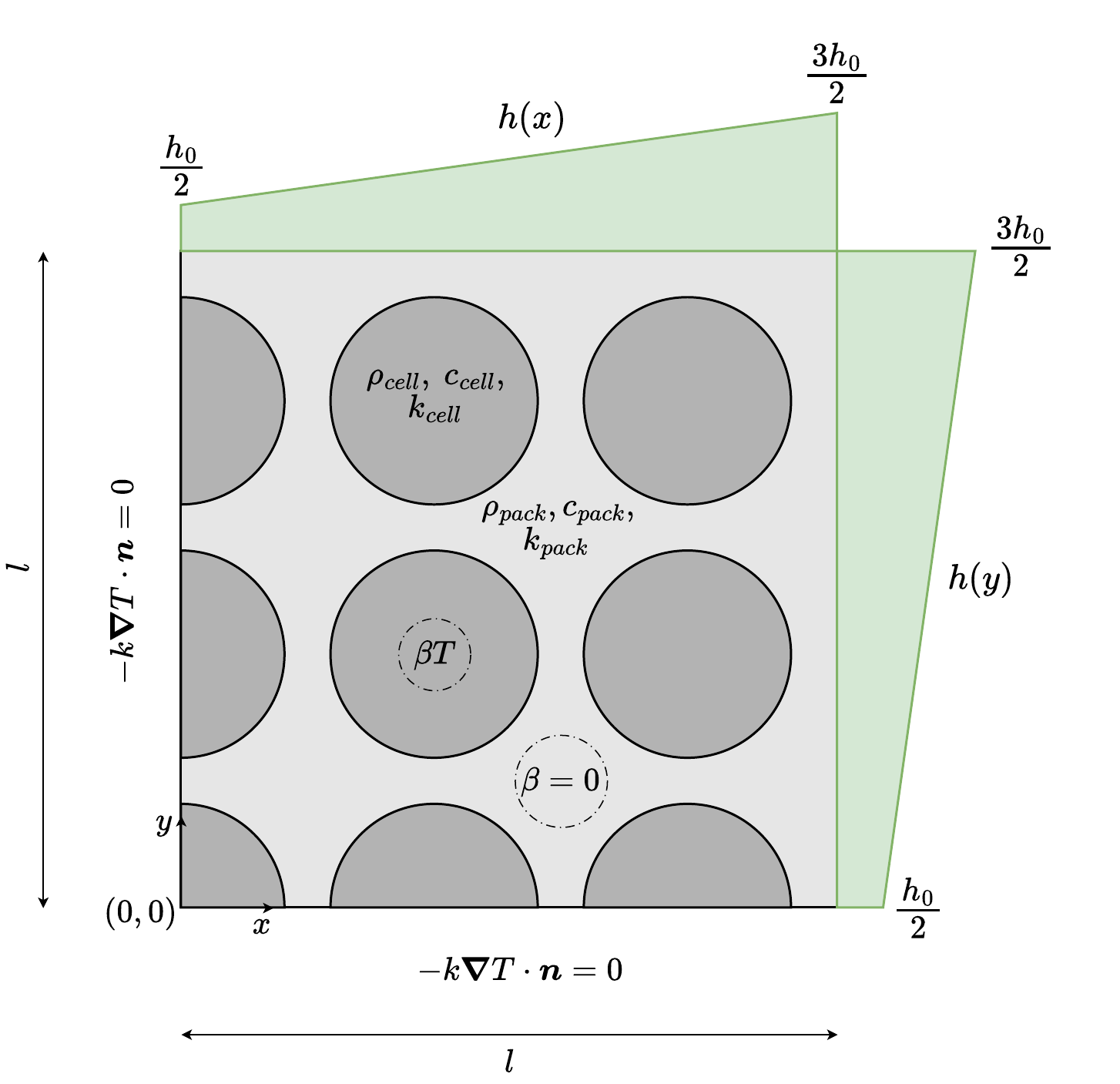}
\caption{Schematic of the quarter battery pack model with spatially varying convective boundary conditions. The heat transfer coefficient is assumed to vary linearly along the external boundaries to represent non-uniform cooling conditions. The variable $h_0$ in the schematic represents the average value of the linear variation. Symmetry boundary conditions are imposed along the internal edges of the quarter domain.}
\label{varhg}
\end{figure}

In the first example problem, the convective heat transfer coefficient is assumed to vary linearly along the convective boundaries, as shown in \autoref{varhg}. Such variations may arise in practical battery packs due to non-uniform airflow or cooling channel design. Stability curves showing the variation of the threshold heat generation parameter ($\beta_{threshold}$) with pack conductivity ($k_{pack}$) are obtained for three cases corresponding to average convection coefficients of $250$, $500$, and $1000 Wm^{-2}K^{-1}$.

For each case, two simulations are performed: one with a constant convection coefficient $h_0$, and the other with a linearly varying coefficient, as shown in \autoref{varhg}, whose spatial average is the same as $h_0$.

\begin{figure}[ht!]
\centering
\includegraphics[trim= {0.0cm 0.0cm 8.0cm 0.0cm},clip,width=0.65\textwidth]{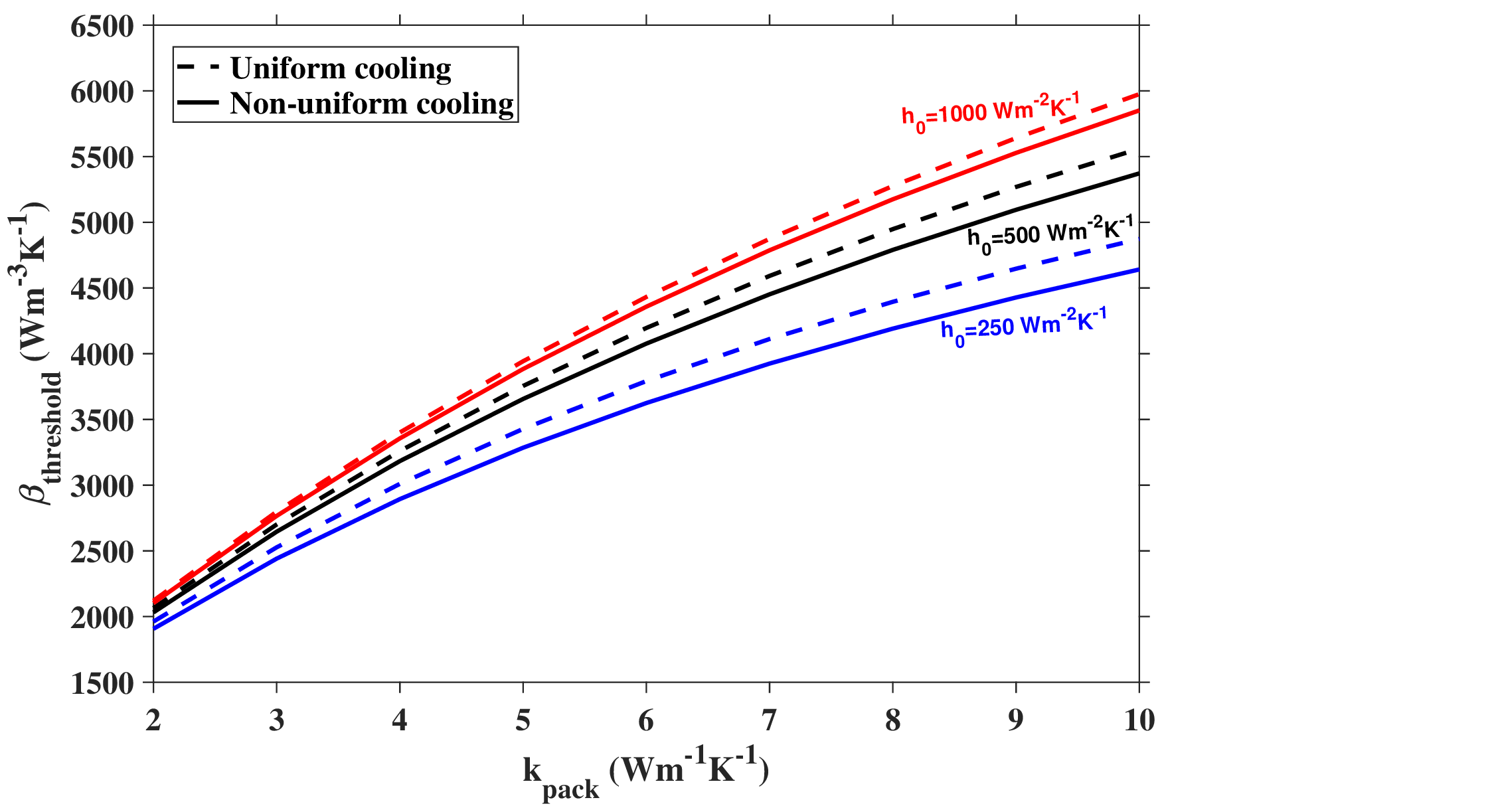}
\caption{Stability curves ($\beta_{threshold}$ vs $k_{pack}$) for the quarter battery pack with spatially varying convection. The plots illustrate the reduction in threshold values of $\beta$ due to non-uniform cooling along the boundaries, compared to uniform cooling.}
\label{varh}
\end{figure}


The resulting stability curves are presented in \autoref{varh}. Results for three different values of $h_0$ consistently indicate that the linearly varying case has lower threshold values of $\beta$ compared to uniform convective cooling. This behavior is attributed to the presence of locally under-cooled regions along the boundary, particularly when the local value of the convective heat transfer coefficient is small. Thermal instability tends to initiate in these poorly cooled regions even when other portions of the boundary experience sufficient heat removal. Once initiated, such thermal instability can propagate throughout the rest of the body. Consequently, non-uniform cooling weakens the overall thermal stability of the pack, resulting in the occurrence of thermal instability at a lower value of $\beta$.

\subsubsection{Spatially varying internal heat generation}

Next, the heat generation parameter $\beta$ within each cell is allowed to vary spatially according to a prescribed function, given by 
\begin{equation}
     \beta(r)=2 \beta_0 \left( 1-\frac{r^2}{R^2} \right),
    \label{f1}
\end{equation}
where $r$ is the distance from the center of the cell and $\beta_0$ represents the average of the prescribed variation. As defined, $\beta(r)$ attains its maximum value at the cell center and gradually decreasing toward the cell boundary. This choice of spatially varying $\beta$ is motivated by the temperature dependence of internal resistance of the cell, and, therefore, heat generation. In general, the core of each cell is expected to be hottest, and, therefore, have a larger value of $\beta$ compared to the outer parts of the cell. In the present problem, the variation in $\beta$ is defined such that the spatial average equals that of the corresponding constant $\beta$ case already considered in prior sections.

\begin{figure}[ht!]
\centering
\includegraphics[trim= {0.0cm 0.0cm 8.0cm 0.0cm},clip,width=0.65\textwidth]{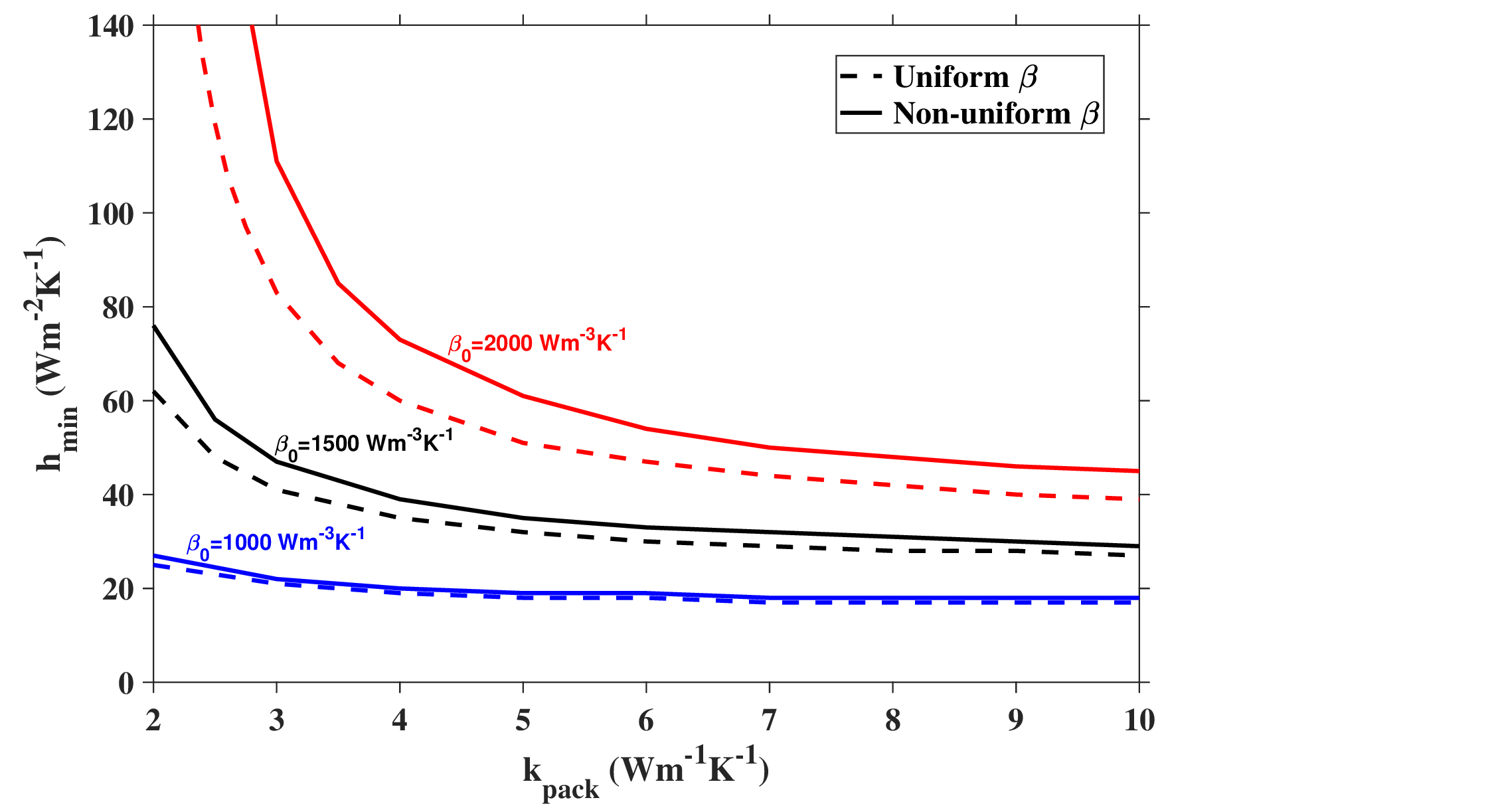}
\caption{Minimum value of uniform convection coefficient, $h_{min}$, required to avoid thermal runaway as a function of pack thermal conductivity $k_{pack}$ for the quarter battery pack, shown for cases of uniform and spatially varying heat generation.}
\label{varb}
\end{figure}

We obtain the stability curves by plotting the minimum value of the uniform convection coefficient $h_{min}$ on all four edges required to prevent thermal runaway as a function of pack thermal conductivity. Curves are presented for the cases of uniform and spatially varying $\beta$ for three values of average heat generation parameter: $\beta_0=1000$, $1500$, and $2000\ Wm^{-3}K^{-1}$ (\autoref{varb}). 



Results show that spatially varying heat generation increases the minimum cooling requirement. Regions near the centers of the cells act as localized hot spots where heat accumulates rapidly, leading to early onset of instability. When the surrounding pack material has low thermal conductivity, the locally generated heat cannot conduct efficiently, further intensifying this effect. However, at higher values of pack thermal conductivity, there is more effective diffusion of heat, and the difference between the constant and varying $\beta$ cases gradually diminishes.

\subsubsection{Influence of cell radius}

\begin{figure}[ht!]
\centering
\includegraphics[trim= {0.0cm 0.0cm 8.0cm 0.0cm},clip,width=0.65\textwidth]{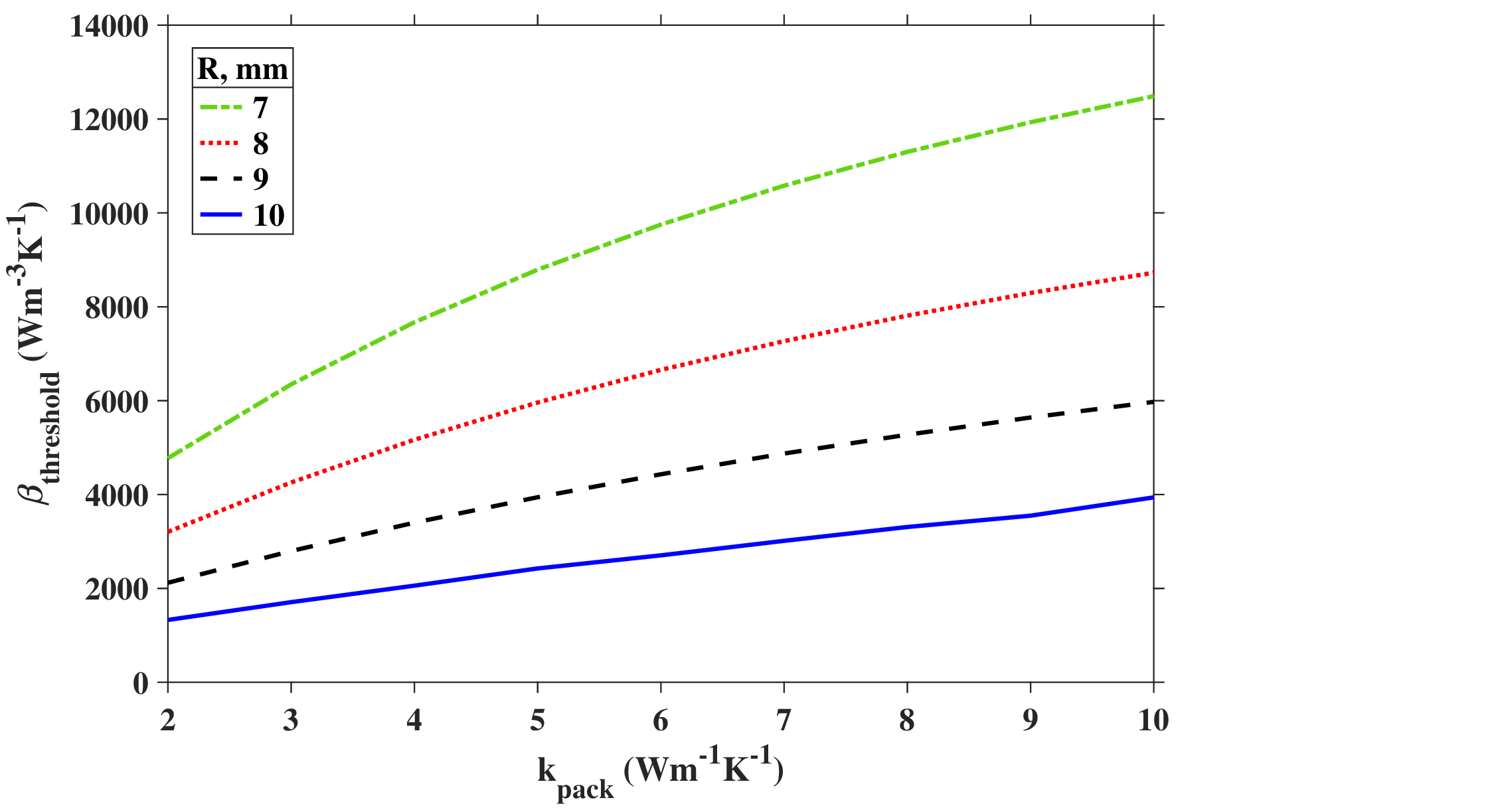}
\caption{Stability curves showing the threshold heat generation parameter $\beta_{threshold}$ versus pack conductivity $k_{pack}$ for different cell radii in the quarter battery pack.}
\label{varr}
\end{figure}

We next investigate the effect of cell size by varying the radius of the cylindrical cells while keeping the overall pack dimensions unchanged. Simulations are carried out for radii of $7$, $8$, $9$, and $10\ mm$, and the corresponding stability curves for a prescribed value of constant $h=1000\ Wm^{-2}K^{-1}$ are shown in \autoref{varr}.

The results reveal a monotonic decrease in the threshold $\beta$ with increasing cell radius. Larger cells generate heat over a greater volume, thereby increasing the total thermal load within the pack. Moreover, as the cell radius increases, the core of the cell becomes farther and farther away from the cooling boundary and thus thermally more isolated. This further increases the propensity for large temperature rise at the core, which may cause the onset of local thermal runaway, and eventually propagation to the rest of the cell and battery pack. Under unchanged cooling conditions, this expanded heat generation region lowers the allowable heat generation rate required to maintain stability. While cell sizing is a considerably complicated exercise, involving several other considerations, such as manufacturability and electrochemical performance, the analysis presented here offers a thermal perspective for cell sizing towards preventing thermal runaway.

\subsubsection{Non-uniform distribution of heat generation across the pack}

\begin{figure}[ht!]
\centering
\includegraphics[trim= {0.0cm 0.0cm 0.0cm 0.0cm},clip,width=0.5\textwidth]{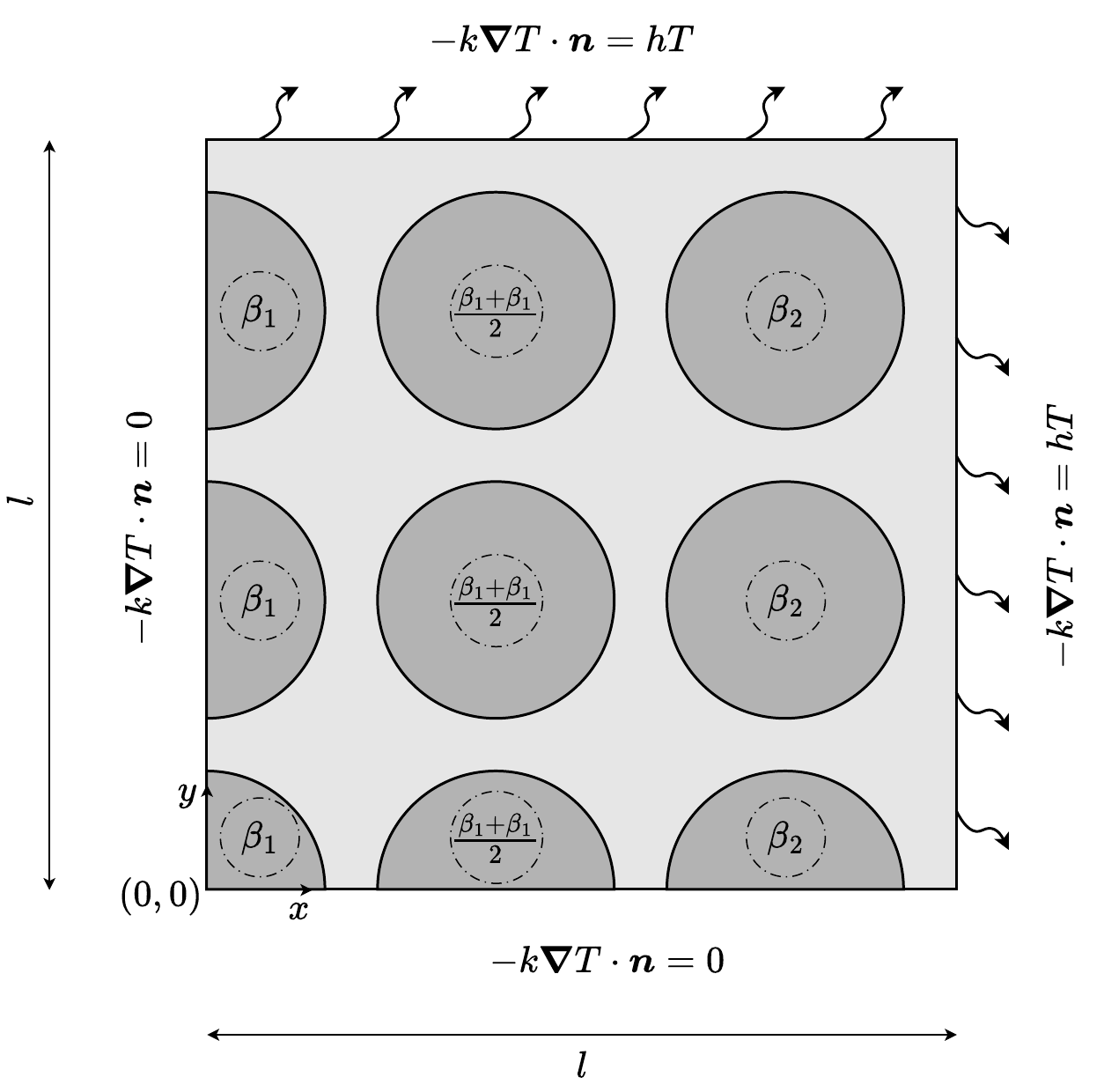}
\caption{Schematic of the quarter battery pack illustrating a varying heat generation distribution, characterized by parameters $\beta_1$ and $\beta_2$ in the cells near the symmetry and convective boundaries, respectively, with linearly interpolated values in the intermediate region.}
\label{varbetadg}
\end{figure}

We examine a scenario in which the distribution of heat generation parameter varies across the pack. Two parameters, $\beta_1$ and $\beta_2$, represent heat generation in cells near the symmetry and convective boundaries, respectively, while the intermediate row is assigned their average, as shown in \autoref{varbetadg}. This produces an approximately linear gradient in heat generation across the pack.

\begin{figure}[ht!]
 \centering
 \includegraphics[trim= {0.0cm 0.0cm 0.0cm 0.0cm},clip,width=0.55\textwidth]{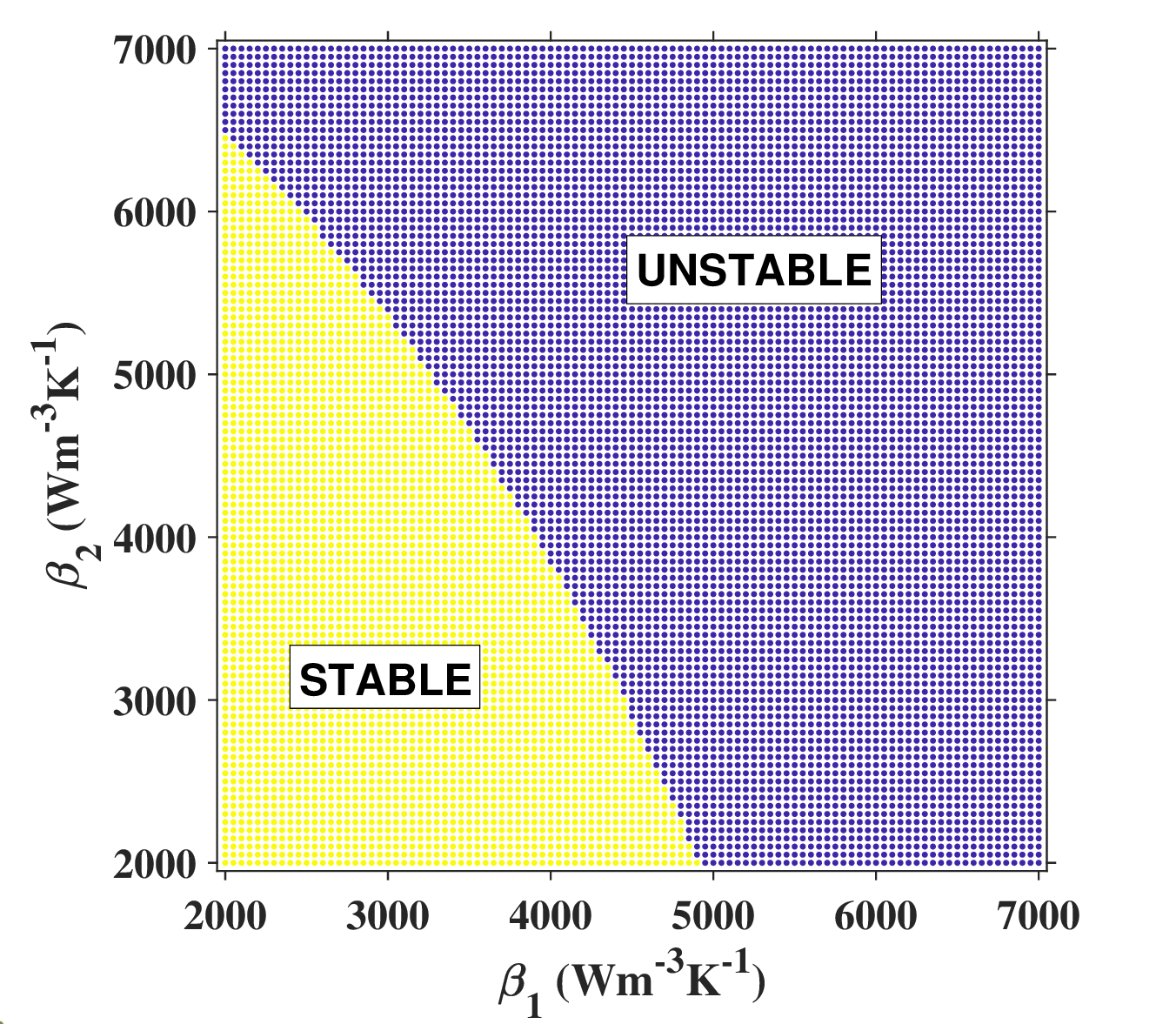}
 \caption{Stability scatter plot in the $\beta_1-\beta_2$ parameter space for the quarter battery pack for a varying distribution of heat generation. The plot identifies safe (stable) and unsafe (unstable) regions, highlighting the asymmetry in stability limits due to higher heat accumulation in the cells and surrounding regions near the adiabatic boundary.}
 \label{b1b2}
 \end{figure}

A scatter plot identifying stable and unstable regions in the $\beta_1$--$\beta_2$ parameter space is presented in \autoref{b1b2}, with both parameters varied between $2000$ and $7000\ Wm^{-3}K^{-1}$.

The results indicate a noticeable asymmetry in stability limits. When $\beta_1$ is minimized, the allowable value of $\beta_2$ is relatively high; however, when $\beta_2$ is minimized, the permissible $\beta_1$ is significantly lower. This behavior reflects greater tendency for heat to accumulate near the symmetry boundary, where cooling is less effective due to greater physical distance from the convective cooling boundaries. As a result, stricter limits on heat generation are required in those regions to maintain overall thermal stability of the battery pack.

While illustrated here for a simplified linear variation, this example highlights the capability of the proposed framework to identify critical heat generation profiles and evaluate worst-case scenarios. The methodology can be readily extended to more complex spatial distributions, making it a valuable tool for thermal design and safety assessment of battery packs.

\subsection{Three-dimensional problems}

\begin{figure}[ht!]
 \centering
     \begin{subfigure}[b]{0.49\textwidth}
         \centering
         \includegraphics[trim= {0.0cm 0.0cm 0.0cm 0.0cm},clip,width=0.65\textwidth]{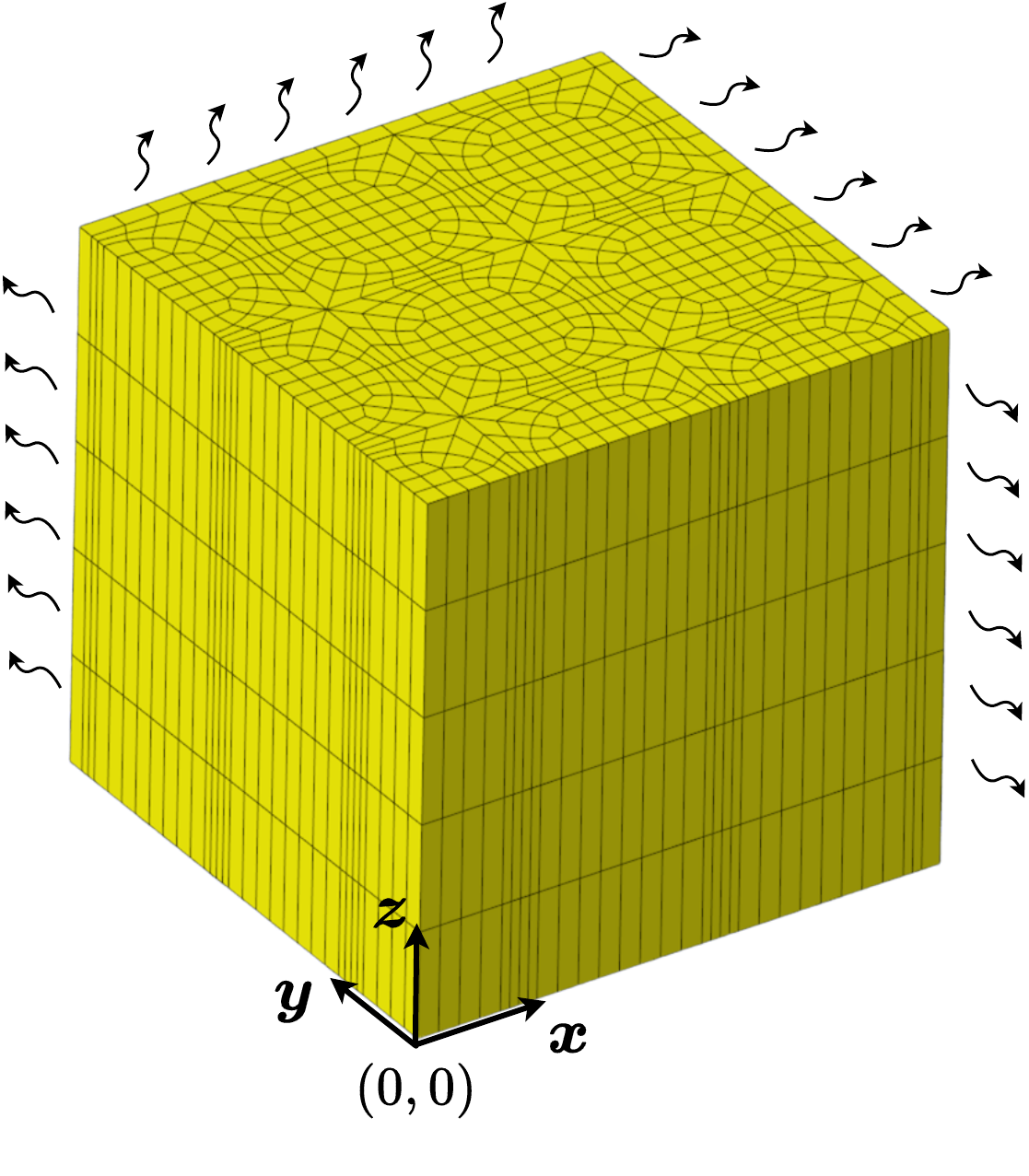}
         \caption{2-CF Configuration.}
         \label{2cf}
     \end{subfigure}
     \begin{subfigure}[b]{0.49\textwidth}
         \centering
         \includegraphics[trim= {0.0cm 0.0cm 0.0cm 0.0cm},clip,width=0.65\textwidth]{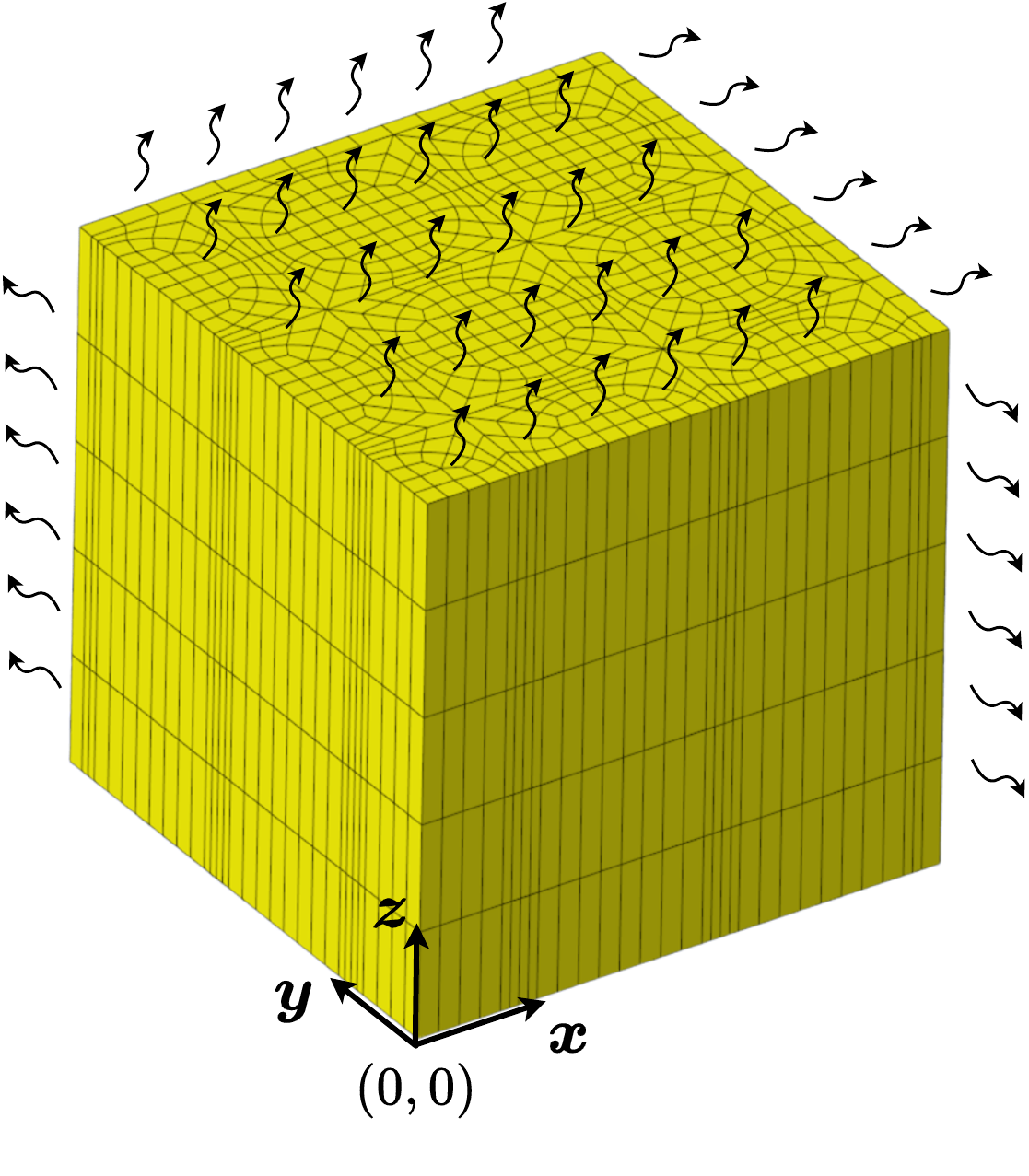}
         \caption{3-CF Configuration.}
         \label{3cf}
     \end{subfigure}
 \caption{\textcolor{black}{Schematic of the three-dimensional battery pack models considered in the present study. \autoref{2cf} and \autoref{3cf} illustrate different cooling conditions: (a) 2 CF configuration and (b) 3 CF configuration, corresponding to cases with two and three convective cooling faces, respectively. Internal heat generation occurs only within the cylindrical cells.
}}
 \label{3dm}
 \end{figure}

We now examine the applicability of the proposed formulation to three-dimensional problems by extending the previously studied battery pack model. The two-dimensional geometry is expanded into the third direction ($z$--direction) to construct a cubical 3D model, as illustrated in \autoref{3dm}. The meshed geometry of the 3D model consists of a total of $3405$ finite elements. Thermophysical properties remain identical to those used in the 2D case.

\begin{figure}[ht!]
\centering
\includegraphics[trim= {0.0cm 0.0cm 8.0cm 0.0cm},clip,width=0.65\textwidth]{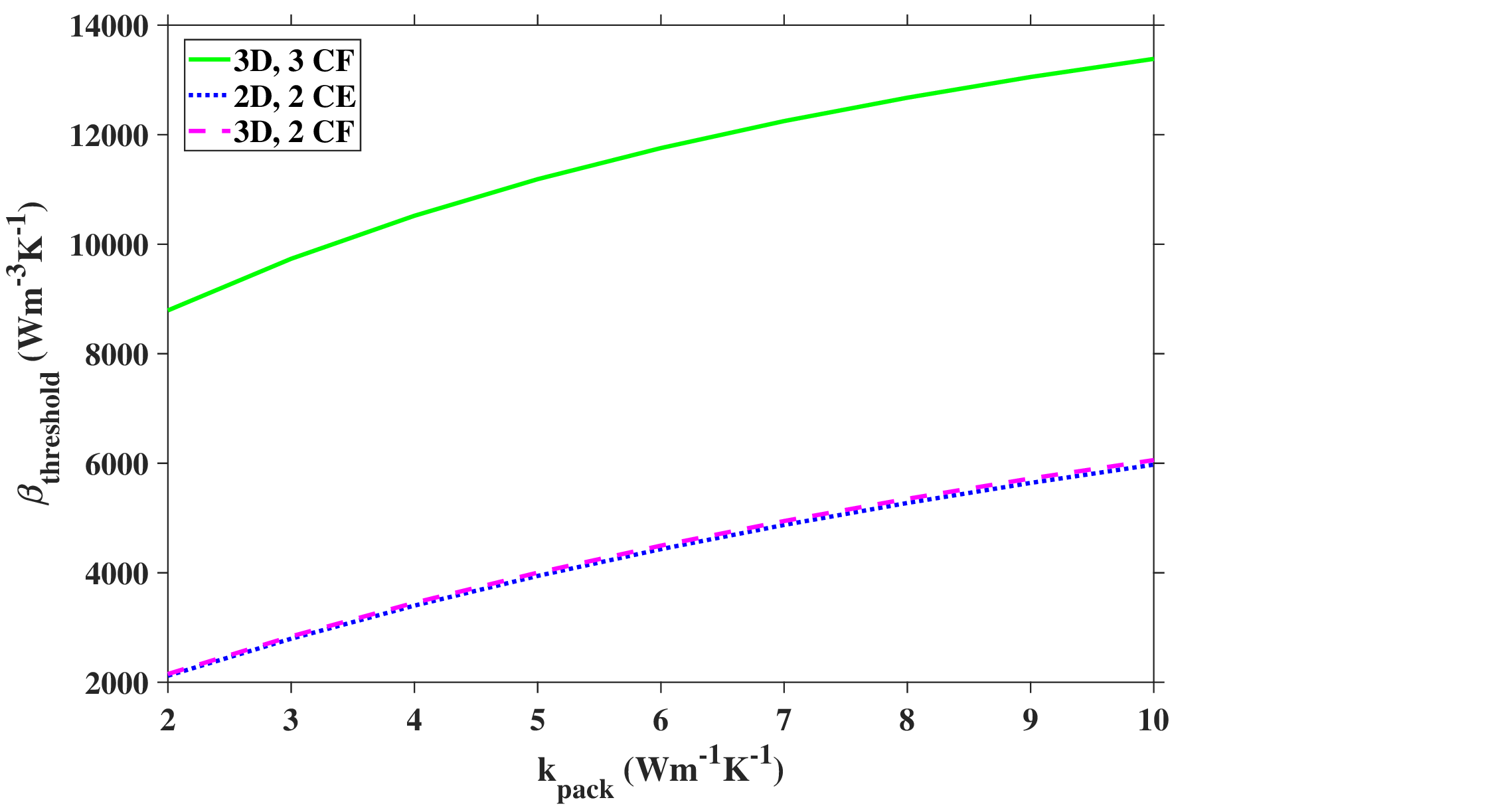}
\caption{Stability curves showing the threshold heat generation parameter ($\beta_{threshold}$) versus pack conductivity ($k_{pack}$) for 2D and 3D battery pack configurations. The three curves correspond to: 2D quarter pack with convection on two edges (2 CE), 3D pack with convection on two faces (2 CF), and 3D pack with convection on three faces (3 CF). The comparison highlights the effect of geometry and convective surface area on thermal stability.}
\label{3d1}
\end{figure}

To ensure a consistent comparison, we prescribe convective boundary conditions identical to those in the 2D configuration. This 3D case is denoted as "2 CF" (i.e., "two convective faces"), indicating that only two faces of the pack are subjected to convection with a uniform heat transfer coefficient $h$, while the remaining faces are treated as adiabatic. Internal heat generation is assumed to occur only within the cylindrical cells with a uniform heat generation parameter $\beta$. We compute the stability curve by evaluating the threshold values of $\beta$ over a range of pack conductivities, $k_{pack} \in [2,10]\ Wm^{-1}K^{-1}$, and compare the results with the corresponding 2D case. The 2D case is denoted as "2 CE" (i.e., "two convective edges"). Owing to similar boundary conditions, the transition from 2D to 3D increases both total heat generation and total heat dissipation in nearly the same proportion. Consequently, only minor deviations are expected in the stability behavior. \autoref{3d1} confirms this expectation, as the stability curves for the 2D and 3D cases nearly coincide, thereby validating the formulation for three-dimensional analyses.

Next, we investigate a modified scenario, denoted as "3 CF" (i.e., "three convective faces"), in which an additional face (the upper surface) is subjected to the same convective boundary condition, as shown in  \autoref{3cf}, representing the full battery pack exposed to convection on all the surfaces. The resulting stability curve, also shown in \autoref{3d1}, exhibits higher threshold values of $\beta$ compared to the two-face convection case. This increase is due to enhanced heat dissipation resulting from the additional convective surface. The increased convective surface area promotes greater heat removal from the system, thereby reducing the effective thermal resistance of the system. As a result, the system remains stable even at higher internal heat generation.

Finally, we consider a spatially varying heat generation parameter within each cell, given by 
\begin{equation}
     \beta(r,z)=3 \beta_0 \left( 1-\frac{r^2}{R^2} \right) \left( 1-\frac{(z-l/2)^2}{(l/2)^2} \right),
    \label{f2}
\end{equation}
\begin{figure}[ht!]
\centering
\includegraphics[trim= {0.0cm 0.0cm 8.0cm 0.0cm},clip,width=0.65\textwidth]{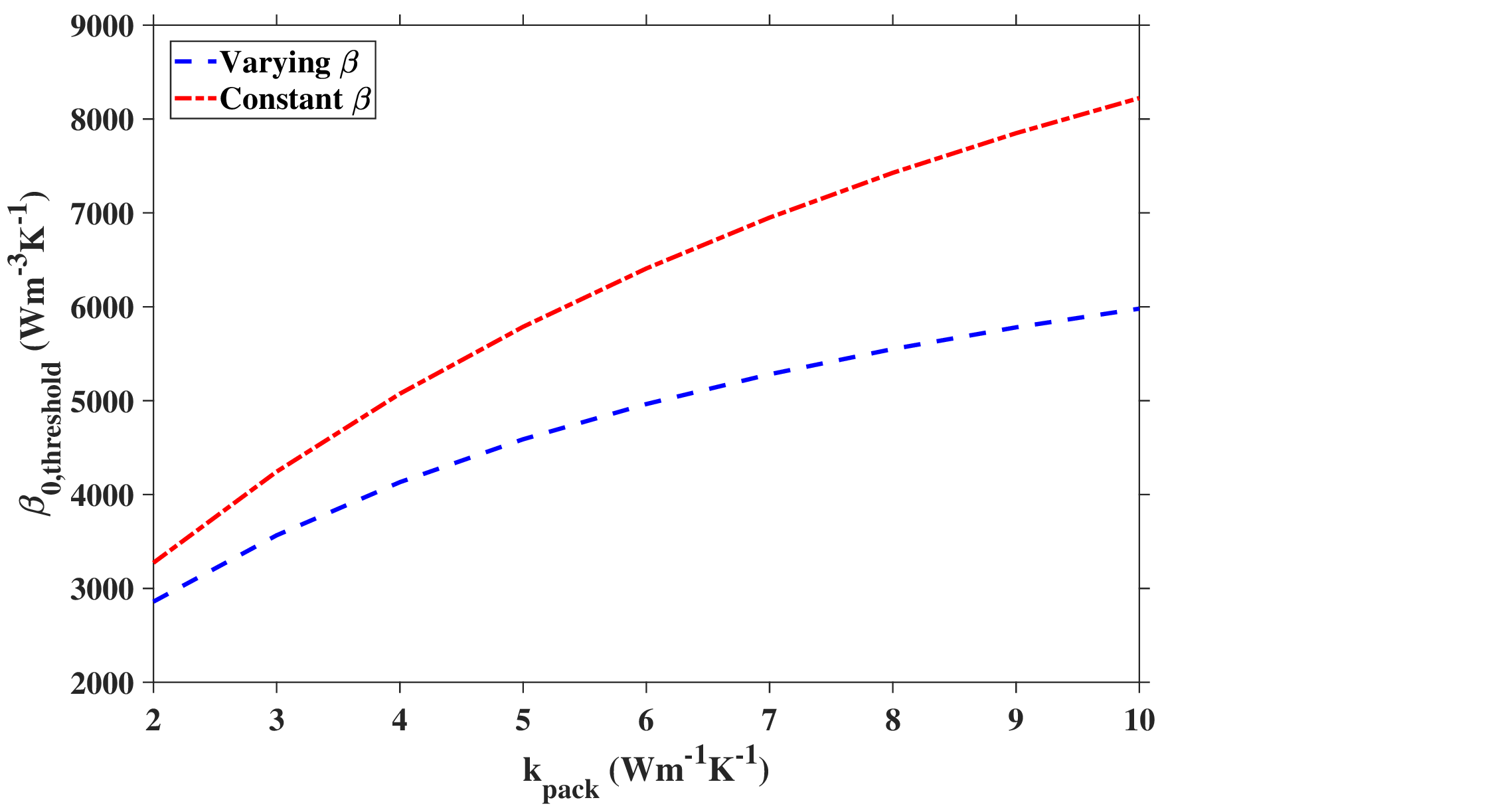}
\caption{Stability curves showing the threshold value of the average heat generation parameter ($\beta_{0,threshold}$) versus pack conductivity ($k_{pack}$) for the 3D battery pack. Two cases are compared: uniform $\beta$ within each cell and spatially varying $\beta$ within each cell with the same average.}
\label{3dv2}
\end{figure}
designed such that the maximum value occurs at the center of each cylindrical cell and gradually decreases toward the boundaries in both radial and axial directions. Here, $r$ is the radial distance from the center of the cell and $\beta_0$ represents the average value of the variation. Two cases are analyzed: one with uniform heat generation and another with spatially varying heat generation while maintaining the same average value. For these cases, the radial and axial thermal conductivities of each cell are taken to be $0.2\ Wm^{-1}K^{-1}$ and $10.0\ Wm^{-1}K^{-1}$, respectively. The stability curves are constructed by plotting the threshold values of the average heat generation parameter against $k_{pack}$ in \autoref{3dv2}. The spatially varying case yields lower threshold values than the uniform case, indicating an earlier onset of thermal instability. This behavior arises from localized thermal hot spots near the cell centers, which trigger instability even when the overall heat generation remains lower than that of the uniform distribution.


Note that while the spatial variation represented by equation \eqref{f2} may appear arbitrary, the goal here is to demonstrate the applicability of the framework developed in this work to a given complicated spatial variation in heat generation. In practical applications, such spatial variation may be obtained from electrochemical analysis of the battery pack.


\section{Conclusions}\label{conclusion}


The key novelty of this work lies in the development of a finite element framework for predicting the onset of thermal runaway in geometrically complicated systems with positive feedback between temperature rise and heat generation, such as Li-ion battery packs. The formulation integrates an eigenvalue-based stability criterion with finite element analysis, making it possible to predict thermal stability of systems with complicated geometries that are difficult to analyze using analytical tools available in literature. Compared to past work, this numerical finite element based approach 
is versatile and flexible in being able to simulate complicated scenarios and geometries that are not possible to analyze using other analytical techniques. The numerical approach developed here provides a clear and computationally efficient means for assessing thermal stability without requiring full transient simulations. The accuracy of the proposed approach has been verified through comparisons with analytical solutions for special cases, demonstrating excellent agreement. The numerical examples presented in the previous section further illustrate the capability of the framework to capture the influence of material properties, boundary conditions, geometric parameters, and spatial variations in heat generation on the stability limits of battery systems. The formulation remains effective and extends naturally to three-dimensional problems that may be more representative of practical battery packs. Overall, the proposed methodology offers a systematic tool for evaluating thermal stability and identifying conditions that may lead to thermal runaway. Extensions of the present framework to  more complex multiphysics settings would be a natural direction for future work. For example, integration with electrochemical modeling to obtain heat generation distribution functions may further increase the relevance of these results to real-life battery packs.



\renewcommand{\bibname}{References}		

\bibliographystyle{elsarticle-num}				
\bibliography{references}	

\end{document}